\documentclass[12pt,letterpaper,english,aps,tightenlinesletterpaper,groupaddress,secnumarabic,nofootinbib]{revtex4}
\usepackage{mathptmx}

\usepackage[T1]{fontenc}
\usepackage[latin9]{inputenc}
\usepackage{xcolor}
\usepackage{babel}
\usepackage{amsmath}
\usepackage{amssymb}
\usepackage{esint}
\PassOptionsToPackage{normalem}{ulem}
\usepackage{ulem}
\usepackage[unicode=true,pdfusetitle,
 bookmarks=true,bookmarksnumbered=false,bookmarksopen=false,
 breaklinks=false,pdfborder={0 0 1},backref=section,colorlinks=false]
 {hyperref}
\hypersetup{
 citecolor=blue}
\usepackage{breakurl}

\makeatletter

\providecolor{lyxadded}{rgb}{0,0,1}
\providecolor{lyxdeleted}{rgb}{1,0,0}

\DeclareRobustCommand{\lyxdeleted}[3]{{\texorpdfstring{\color{lyxdeleted}\sout{#3}}{}}}

\@ifundefined{textcolor}{}
{%
 \definecolor{BLACK}{gray}{0}
 \definecolor{WHITE}{gray}{1}
 \definecolor{RED}{rgb}{1,0,0}
 \definecolor{GREEN}{rgb}{0,1,0}
 \definecolor{BLUE}{rgb}{0,0,1}
 \definecolor{CYAN}{cmyk}{1,0,0,0}
 \definecolor{MAGENTA}{cmyk}{0,1,0,0}
 \definecolor{YELLOW}{cmyk}{0,0,1,0}
}

\usepackage{babel}
\usepackage{babel}
\usepackage{tensind}

\makeatother

\begin{document}

\title{Lorentz invariance of basis tensor gauge theory}

\author{Edward Basso}

\email{ebasso@wisc.edu}

\affiliation{Department of Physics, University of Wisconsin-Madison, Madison,
WI 53706, USA}

\author{Daniel J. H. Chung}

\email{danielchung@wisc.edu}

\affiliation{Department of Physics, University of Wisconsin-Madison, Madison,
WI 53706, USA}
\begin{abstract}
Basis tensor gauge theory (BTGT) is a vierbein analog reformulation
of ordinary gauge theories in which the vierbein field describes the
Wilson line. After a brief review of the BTGT, we clarify the Lorentz
group representation properties associated with the variables used
for its quantization. In particular, we show that starting from an
SO(1,3) representation satisfying the Lorentz-invariant U(1,3) matrix
constraints, BTGT introduces a Lorentz frame choice to pick the Abelian
group manifold generated by the Cartan subalgebra of u(1,3) for the
convenience of quantization even though the theory is frame independent.
This freedom to choose a frame can be viewed as an additional symmetry
of BTGT that was not emphasized before. We then show how an $S_{4}$ permutation symmetry and a parity symmetry of frame fields
natural in BTGT can be used to construct renormalizable gauge theories
that introduce frame dependent fields but remain frame independent
perturbatively without any explicit reference to the usual gauge field.
\end{abstract}
\maketitle

\section{Introduction}

Rewriting gauge theories in novel formalisms continue to offer insights
into both computational techniques and ideas for physics beyond the
SM (see e.g.~\cite{Arkani-Hamed:2013jha,Arkani-Hamed:2017jhn,Badger:2005zh,Elvang:2013cua,Henn:2014yza,Christensen:2018zcq,Witten:1998qj,Aharony:1999ti}).
In analogy with general relativity, ordinary gauge theories of semisimple
compact Lie groups (see e.g.~\cite{Yang:1954ek,Abers:1973qs,Itzykson:1980rh,Polyakov:1987ez,Sterman:1994ce,Hooft:1995gh,Weinberg:1996kr})
can be rewritten in terms of vierbeins of the gauge group space. The
quantization of this vierbein theory was called basis tensor gauge
theory (BTGT) \cite{Chung:2016lhv}. Previous works \cite{Chung:2016lhv,Chung:2017zck,Basso:2019yap}
focused on a symmetric Lorentz group representation $G_{(f)}^{\mu\nu}$
of the vierbein field, which transforms as 
\begin{equation}
\left[G_{(f)\,\beta}^{\alpha}(x)\right]^{i}\rightarrow\left[G_{(f)\,\beta}^{\alpha}(x)\right]^{j}\left[g^{-1}\left(x\right)\right]^{ji}
\end{equation}
where $g$ is the representation of the ordinary gauge group. The
path integral quantization of $G_{(f)}^{\mu\nu}$ was accomplished
using a field redefinition to $\theta_{a}^{B}(x)$ phase variables
at the expense of introducing a global field $(H^{a})^{\mu\nu}$ which
transforms as a tensor under Lorentz transformations. The Lorentz
group representation theory interpretation of $(H^{a})^{\mu\nu}$
is somewhat obscure and the underlying reasons why the global field
introduction does not lead to pathologies have not been addressed
previously.

In this work, we therefore clarify the Lorentz group representation
of $G_{(f)}^{\mu\nu}$ and its associated representational meaning
of this global field $(H^{a})_{\mu\nu}$. We show that $G_{(f)}^{\mu\nu}$
is a set of symmetric $SO(1,3)$ complex tensors (closed under the
Lorentz transformation orbit before restricting to the $\theta_{a}^{B}(x)$
functional space) satisfying the Lorentz invariant $U(1,3)$ matrix
constraints. The set of four $\theta_{a}^{B}$ (for a fixed index
$B$) parameterizes the $U(1)^{4}$ generated by the Cartan subalgebra
of $u(1,3)$. More importantly, we show that $(H^{a})_{\mu\nu}$ is
a complete set of frame-dependent projections of the usual Lorentz
invariant metric $\eta_{\mu\nu}$ which is why there are no global
field pathologies. The naive dangers of frame dependence (partly arising
from the non-linearity of $G_{(f)}^{\mu\nu}$ map to $\theta_{a}$)
is argued to be manifestly innocuous because of the reliance of BTGT
on the ordinary gauge field $A_{\mu}$ in defining the path integral.
Previously stated defining symmetry of the quantized theory (gauge
and BTGT invariance) is extended to include frame independence of
different choices of $(H^{a})_{\mu\nu}$. Finally, we give one recipe
for constructing frame independent gauge theories using frame dependent
variables based on frame covariance, gauge invariance, BTGT invariance,
$S_{4}$ permutation symmetry, and a parity symmetry, without making
any explicit reference to the usual gauge field $A_{\mu}$.

The order of presentation is as follows. In Sec.~\ref{sec:A-brief-review},
we give a brief review of the BTGT theory. In Sec.~\ref{sec:What-is-?whatisH},
we explain how $(H^{a})_{\mu\nu}$ is equivalent to $\eta_{\mu\nu}$
and explain how the frame dependent description of frame independent
theories arise. An interesting idea in this section is the effect
of the non-linearity of the field redefinition (of in going from the
$G_{(f)}^{\mu\nu}$ description to $\theta_{a}$ description) on the
loss of manifest frame independence. In Sec.~\ref{sec:The-covariant-degrees},
we explain how $G_{(f)}^{\mu\nu}$ forms a set of $SO(1,3)$ tensors
satisfying the Lorentz invariant $U(1,3)$ matrix constraints while
$\theta_{a}^{B}$ is related to the $U(1)^{4}$ subgroup generated
by the Cartan subalgebra of $u(1,3)$. We further explain how the
previous symmetries defining the quantized BTGT theory is extended
to include frame independence. In Sec.~\ref{sec:Frame-independence-without}
we present a theorem illustrating how one can construct frame independent
gauge theories based on frame dependent tensors, using gauge symmetry,
BTGT symmetry, $S_{4}$ symmetry, and a parity symmetry, without any
explicit reference to the gauge field $A_{\mu}$.

By convention, all repeated indices will be summed unless noted otherwise
or is clear from the context of the two sides of the equation. This
notational issue should be kept in mind because parts of the paper
will explain the relationship between the non-manifestly-covariant
representations and covariant representations, and this summation
convention applies to noncovariant indices as well.

\section{\label{sec:A-brief-review}A brief review of BTGT}

Let's briefly review the BTGT theory. For an explanation of many statements
made in this section, see Refs.~\cite{Basso:2019yap,Chung:2016lhv,Chung:2017zck}.

Suppose $\phi$ is a matter field which has the following gauge transformation
property 
\begin{equation}
\phi^{k}(x)\rightarrow\left[g(x)\right]^{ks}\phi^{s}(x)
\end{equation}
\begin{equation}
\left[g(x)\right]^{ks}\equiv\left(e^{i\Gamma^{C}(x)T^{C}}\right)^{ks}
\end{equation}
where $T^{C}$ are the Hermitian generators satisfying the usual Lie
algebraic relationship $[T^{A},T^{B}]=if^{ABC}T^{C}$. The covariant
derivative associated with this matter field has a connection, usually
expressed as a Lorentz vector field, which tells how to define parallel
transports keeping the ``gauge direction'' parallel as the field
is transported along a curve. An alternate formulation of this type
of mathematics, typically employed in general relativity to import
the spinor technology, involves defining a set of local direction
fields having tensor indices in gauge space as well as Minkowski space
with which one can use to convert directions in the gauge space into
directions in Minkowski space. With these direction fields, called
vierbein fields, one can reproduce the information contained in the
original connection field commonly denoted as $A_{\mu}^{C}T^{C}$.

One such choice for the gauge vierbeins is $G_{(f)\,\beta}^{\alpha}(x)$,
which is a complex Lorentz tensor field that transforms as an $\bar{R}$
from the right under the gauge group representation and as a rank
2 Lorentz projection tensor. It has the properties of a gauge vierbein
in that 
\begin{equation}
\left[G_{(f)\,\beta}^{\alpha}(x)\right]^{i}\rightarrow\left[G_{(f)\,\beta}^{\alpha}(x)\right]^{j}\left[g^{-1}\left(x\right)\right]^{ji},\label{eq:Gtransform-1}
\end{equation}
and it essentially maps the gauge space to a Lorentz tensor space,
at least locally. The analogy with gravitational vierbeins $(e_{a})_{\mu}$
is the following: the indices $\{f,\alpha,\beta\}$ are the analogs
of the fictitious Minkowski space index $a$ of $(e_{a})_{\mu}$,
and the representation of Eq.~(\ref{eq:Gtransform-1}) is the analog
of the diffeomorphism acting on the $\mu$ index of ($e_{a})_{\mu}$.
Explicitly, it allows one to express the gauge field as 
\begin{equation}
A_{\mu}=i\left[G^{-1\alpha\beta}\right]\left[\partial_{\alpha}G_{\beta\mu}\right]\label{eq:agrelation-1}
\end{equation}
where $G_{\beta\mu}$ are related to the basis tensor as 
\begin{equation}
\left[G_{\beta\mu}\right]^{qm}=\sum_{f}^{{\rm dim}R}\xi_{(f)}^{q}\left[G_{(f)\beta\mu}\right]^{m}\label{eq:Gmunudef}
\end{equation}
where $\xi_{(f)}^{k}$ are constant vectors that span the gauge group
representation $R$ space as 
\begin{equation}
\delta^{kl}=\sum_{f}^{{\rm dim}R}\xi_{(f)}^{k}\xi_{(f)}^{*l}.\label{eq:normalbasis-1}
\end{equation}
Quite interestingly and suggestively, Eq.~(\ref{eq:agrelation-1})
is in the form of a sigma model.

Now, we introduce the variables that will be used to pose the question
of this paper. To quantize the theory, previous works \cite{Basso:2019yap,Chung:2016lhv,Chung:2017zck}
used the following representation involving a symmetric tensor $\left[G_{\beta\mu}\right]^{qm}$:
\begin{equation}
\left(\left[G_{(f)}(x)\right]_{\phantom{\gamma}\delta}^{\gamma}\right)^{j}=\xi_{(f)}^{*l}\left[\left(\exp\left[-i\sum_{a=0}^{3}\theta_{a}^{M}(x)H^{a}T^{M}\right]\right)_{\phantom{\gamma}\delta}^{\gamma}\right]^{lj}
\end{equation}
or equivalently 
\begin{equation}
G_{\phantom{\gamma}\delta}^{\gamma}=\left(\exp\left[-i\sum_{a=0}^{3}\theta_{a}^{M}(x)H^{a}T^{M}\right]\right)_{\phantom{\gamma}\delta}^{\gamma}\label{eq:exponentialmap}
\end{equation}
where $(H^{a})^{\mu\nu}$ is a global real field which transforms
as a symmetric Lorentz tensor and $\theta_{a}^{M}(x)$ is a real Lorentz
scalar field. In practice, one can choose $H^{a}$ explicitly as 
\begin{equation}
(H^{a})^{\mu\nu}=\sum_{b}\psi_{(a)}^{\mu}\psi_{(b)}^{\nu}\eta^{ab}=\frac{\psi_{(a)}^{\mu}\psi_{(a)}^{\nu}}{\psi_{(a)}\cdot\psi_{(a)}}\label{eq:intermsofpsi}
\end{equation}
where $\psi_{(a)}^{\mu}$ are four real 4-vectors satisfying $\psi_{(a)}^{\mu}\psi_{(b)\mu}=\eta_{ab}$,
where $\eta_{ab}$ coincides with the components of the Lorentzian
metric in Cartesian coordinates. This choice of $\psi_{(a)}^{\mu}$
normalization will be generalized later (see Appendix \ref{sec:Definitions})
but can always be made without loss of generality.

The introduction of variables of Eq.~(\ref{eq:exponentialmap}) was
particularly useful for the Abelian theory, where one can set $T^{M}$
to a real number and $\xi_{(f)}=1$ because the map between the gauge
field and $\theta_{a}$ simplifies to 
\begin{equation}
A_{\mu}=\sum_{a}(H^{a})_{\phantom{\lambda}\mu}^{\lambda}\partial_{\lambda}\theta_{a},\label{eq:Arelationship}
\end{equation}
which is linear. In the non-Abelian case, the relationship is nonlinear:
\begin{equation}
A_{\mu}=i\sum_{a}U_{a}(H^{a})_{\phantom{\lambda}\mu}^{\lambda}\partial_{\lambda}U_{a}^{\dagger}.\label{eq:nonabelianrelationship}
\end{equation}
where 
\begin{equation}
U_{a}\equiv\exp\left[i\theta_{a}^{A}T^{A}\right].\label{eq:uvariable-1}
\end{equation}
The gauge transformation on $\theta_{a}^{A}$ can be written as 
\begin{equation}
U_{a}\rightarrow e^{i\Gamma}U_{a}\label{eq:gaugetransform-1}
\end{equation}
where $\Gamma\equiv\Gamma^{B}T^{B}.$ Explicitly, Baker\textendash Campbell\textendash Hausdorff
formula gives 
\begin{equation}
\theta_{a}'=\Gamma+\theta_{a}+\frac{i}{2}[\Gamma,\theta_{a}]-\frac{1}{12}\left([\Gamma,[\Gamma,\theta_{a}]]+[\theta_{a},[\theta_{a},\Gamma]]\right)+...
\end{equation}
where $\theta_{a}\equiv\theta_{a}^{A}T^{A}$.

One of the most important new features of a theory with $\theta_{a}$
is that to reproduce ordinary gauge theories, there needs to be a
new \emph{continuous} symmetry which can be referred to as BTGT symmetry\footnote{It is not known whether other symmetries that have the same desired
effect exists. Indeed, one way to interpret this paper is to clarify
and add to this symmetry, as we explain later.}: 
\begin{equation}
U_{a}\rightarrow U_{a}e^{iZ_{a}}\label{eq:BTGT sym for Ua}
\end{equation}
where $Z_{a}\equiv T^{B}Z_{a}^{B}$ satisfies 
\begin{equation}
(H^{a})_{\phantom{\lambda}\mu}^{\lambda}\partial_{\lambda}Z_{a}=0\label{eq:Zconstraint}
\end{equation}
and $U_{a}$ is defined in Eq.~(\ref{eq:uvariable-1}). Note that
even though this is a continuous symmetry similar to a gauge symmetry,
it is a symmetry without any compensating fields transforming as in
gauge transforms. Its main purpose is to package $\theta_{a}^{B}$
into $A_{\mu}$ without any other observables that depend on $\theta_{a}^{B}$
alone. Without this symmetry (or something similar), there would be
the usual problems associated with higher derivative theories stemming
from Eq.~(\ref{eq:Arelationship}) (with a finite power truncation
in higher derivatives) and/or global gauge charge violations. It is
interesting to note that this BTGT symmetry can be expressed in terms
of a transformation of $G_{\phantom{\mu}\nu}^{\mu}$ independently
of the decomposition in $H^{a}$. However, we will defer the exploration
of this topic to a future work.

If one solves Eq.~(\ref{eq:nonabelianrelationship}) for $\theta_{a}^{A}$,
it will depend on the boundary conditions. Eq.~(\ref{eq:BTGT sym for Ua})
can be viewed as defining an equivalence relation $\theta_{a}^{A(1)}\sim\theta_{a}^{A(2)}$
where the $(1)$ and $(2)$ superscripts specify different boundary
conditions to Eq.~(\ref{eq:Arelationship}). Indeed, instead of defining
specific transformations such as Eq.~(\ref{eq:BTGT sym for Ua}),
one can define a generalized BTGT invariance to be the set of transformations
that leave Eq.~(\ref{eq:nonabelianrelationship}) invariant. Given
a curve parameterized by 
\begin{equation}
x_{(c)}^{\mu}(s)\equiv s\psi_{(c)}^{\mu}+x_{0}^{\mu},
\end{equation}
one can integrate Eq.~(\ref{eq:nonabelianrelationship}) to obtain
\begin{equation}
U_{c}^{\dagger}(x_{(c)}(s))=U_{c}^{\dagger}(x_{(c)}(s_{i}))\bar{P}\left[\exp\left(-i\int_{s_{i}}^{s}d\tau A_{\mu}(x_{(c)}(\tau))\psi_{(c)}^{\mu}\right)\right],
\end{equation}
where $\bar{P}$ denotes anti-path-ordering. Exponential of $\theta_{a}$
is clearly is related to the Wilson line.

The action for the BTGT theory is of the form 
\begin{equation}
\mathcal{L}=\frac{-1}{4g^{2}T(R)}{\rm Tr}\left(F^{\mu\nu}F_{\mu\nu}\right),\label{eq:puregluon-1}
\end{equation}
where 
\begin{equation}
-iF_{\mu\nu}=\partial_{\mu}\left[\left(G^{-1}\right)_{\phantom{\lambda}\kappa}^{\lambda}\partial_{\lambda}G_{\phantom{\alpha}\nu}^{\kappa}\right]-\partial_{\nu}\left[\left(G^{-1}\right)_{\phantom{\lambda}\kappa}^{\lambda}\partial_{\lambda}G_{\phantom{\alpha}\mu}^{\kappa}\right]+[\left(G^{-1}\right)_{\phantom{\lambda}\kappa}^{\lambda}\partial_{\lambda}G_{\phantom{\lambda}\mu}^{\kappa},\left(G^{-1}\right)_{\phantom{\lambda}\alpha}^{\rho}\partial_{\rho}G_{\phantom{\alpha}\nu}^{\alpha}].\label{eq:manifest}
\end{equation}
In terms of $\theta_{a}$ variables (embedded into $U_{a}$), this
is 
\begin{eqnarray}
-iF_{\mu\nu} & = & \partial_{\mu}(\sum_{a=0}^{3}U_{a}(H^{a})_{\phantom{\lambda}\nu}^{\lambda}\partial_{\lambda}U_{a}^{\dagger})-\partial_{\nu}(\sum_{a=0}^{3}U_{a}(H^{a})_{\phantom{\lambda}\mu}^{\lambda}\partial_{\lambda}U_{a}^{\dagger})\nonumber \\
 &  & +[\sum_{a=0}^{3}U_{a}(H^{a})_{\phantom{\lambda}\mu}^{\lambda}\partial_{\lambda}U_{a}^{\dagger},\sum_{b=0}^{3}U_{b}(H^{b})_{\phantom{\lambda}\nu}^{\rho}\partial_{\rho}U_{b}^{\dagger}]\label{eq:nonmanifest}
\end{eqnarray}
where we have used 
\begin{equation}
G_{\phantom{\lambda}\nu}^{\lambda}=\sum_{a=0}^{3}(H^{a})_{\phantom{\lambda}\nu}^{\lambda}U_{a}^{\dagger}.\label{eq:btgtfield}
\end{equation}
The introduction of the $\theta_{a}$ variables (through $U_{a}$
and $H^{a}$) is useful for path integral quantization.

In the next section, we explain one of the main points of this paper
which is to explain how the theory does not have a preferred frame
despite the appearance of $(H^{a})_{\phantom{\lambda}\mu}^{\lambda}$.

\section{\label{sec:What-is-?whatisH}What is $(H^{a})^{\lambda\mu}$?}

One obvious question arises as to the interpretation of $(H^{a})^{\lambda\mu}$
transforming as a symmetric Lorentz tensor. For example, one might
very naively guess that $\theta_{a}^{A}(H^{a})^{\lambda\mu}$ which
enters as a package indicates that it is similar to how $\bar{\Psi}\gamma^{\mu}\Psi$
in spinor representation (say for a Dirac field $\Psi$) transforms
as a Lorentz vector. Recall that in the case of $\bar{\Psi}\gamma^{\mu}\Psi$,
the field $\Psi$ transforms as a spinor inducing the behavior as
if $\gamma^{\mu}$ transforms. This means, we can always keep $\gamma^{\mu}$
to be the same set of numbers in any Lorentz frame. Indeed, $\Psi$
field has 4 real functional degrees of freedom to encode the Lorentz
4-vector representation.However, a symmetric Lorentz tensor field
that is not traceless has 10 functional degrees of freedom, whereas
$\theta_{a}^{B}(H^{a})^{\lambda\mu}$ for a fixed $B$ has only 4
functional degrees of freedom. Hence, the object $(H^{a})^{\lambda\mu}$
must transform under Lorentz transformations independently of $\theta_{a}^{B}$.
This means that the situation is not analogous to the spinor representation
and $\theta_{a}^{B}$ degrees of freedom are frame dependent fields.
We clarify the precise covariant representational meaning of $(H^{a})^{\lambda\mu}$
in this section. 

To see what $(H^{a})^{\mu\nu}$ means, first note that it can be written
in terms of $\psi_{(a)}^{\mu}$ according to Eq.~(\ref{eq:intermsofpsi})
which means that its 10 global degrees of freedom are in the constrained
set of real 4-vectors $\psi_{(a)}^{\mu}$. We will now show that $\psi_{(a)}^{\mu}$
is just a set of frame-choice induced projections of ordinary tensors
used in constructing Lorentz invariant field theories.

In ordinary field theory description, a global field such as $\psi_{(a)}^{\mu}$
naively seems exotic. However, for every choice of coordinates $x^{\mu}[n]$
(where ``$[n]$'' labels a particular coordinate system), one implicitly
defines in ordinary field theory the following global field 
\begin{equation}
\Delta=\Delta_{\phantom{\mu}\lambda}^{\mu}e_{\mu}[n]\otimes e^{\lambda}[n]\equiv\delta_{\phantom{\mu}\lambda}^{\mu}e_{\mu}[n]\otimes e^{\lambda}[n]\label{eq:defofbracket}
\end{equation}
whose components are Lorentz invariant (although dependent on $e^{\mu}[n]$
in general). In the BTGT theory with symmetric gauge vierbeins $G^{\mu\nu}$
defined in Eq.~(\ref{eq:Gmunudef}), we first introduce a coordinate
frame ``$[1]$'' which can be chosen arbitrarily since the vacuum
($\theta_{a}=\Gamma$) is Lorentz invariant. We then define $\psi_{(a)}^{\mu}$
as the following projection of the $\Delta$ tensor: 
\begin{equation}
\psi_{(a)}^{\mu}e_{\mu}[n]\equiv\Delta_{\phantom{\mu}\lambda}^{\mu}e_{\mu}[n]\otimes e^{\lambda}[n](e_{a}[1])\label{eq:defofglobalfield}
\end{equation}
where $e_{a}[1]$ refers to an $a$th basis object of an arbitrarily
chosen frame ``$[1]$'' whose coordinate basis is spanned by $e_{\mu}[1]$
and the ``$(...)$'' denotes the usual dual space defined projection
\begin{equation}
e^{\lambda}[1](e_{\mu}[1])=\delta_{\phantom{\lambda}\mu}^{\lambda}.
\end{equation}
This means that as long as we choose \textbf{\uline{4 independent
objects}} $e_{a}[1]$, $\psi_{(a)}^{\mu}$ can be used to construct
$\Delta_{\phantom{\mu}\lambda}^{\mu}$ which is obviously Lorentz
invariant. As we noted, $\Delta$ is used in ordinary field theory.
Only difference between an ordinary field theory and the one with
$\psi_{(a)}^{\mu}$ is the arbitrary initial frame choice ``$[1]$''
and the fact that one \textbf{can} (but do not have to) now describe
field effects that prefer a frame ``$[1]$''.

For an explicit coordinate dependent expression for $\psi_{(a)}^{\mu}$,
consider any frame ``$[2]$'' related to frame ``{[}1{]}'' by
a Lorentz transform 
\begin{equation}
e^{\lambda}[2]=\Lambda_{\phantom{\lambda}\kappa}^{\lambda}e^{\kappa}[1].
\end{equation}
We can evaluate Eq.~(\ref{eq:defofglobalfield}) as 
\begin{equation}
\psi_{(a)}^{\mu}e_{\mu}[2]=\Lambda_{\phantom{\lambda}a}^{\mu}e_{\mu}[2]\label{eq:definingproperty}
\end{equation}
which is the usual basis we used in the BTGT papers. What condition
should we place on $\psi_{(a)}^{\mu}$ usage\textbf{ }in the path
integral to ensure that the theory does not depend on that initial
arbitrary frame $x^{\mu}[1]$ used to define $\psi_{(a)}^{\mu}$?
It is clear that the answer is that \emph{the path integral should
be rewritable without $\psi_{(a)}^{\mu}$ (and its ``a''-index associated
tensors) }\textbf{\emph{and}}\emph{ maintain manifest covariance in
having a manifest Lorentz scalar action and have a Lorentz invariant
path integral measure}. If that were not true, then the physical system
would have an observable that probes the properties endowed by the
$\psi_{(a)}^{\mu}$ which would mean that an experiment can be done
to pick out the arbitrary ``{[}1{]}'' frame with which $\psi_{(a)}^{\mu}$
was defined. We will call this \textbf{frame independence}.

If the local field is a Lorentz tensor whose \emph{definition} is
\emph{independent} of $\psi_{(a)}^{\mu}$, then $\psi_{(a)}^{\mu}$
should appear in combinations such that it sums to invariant tensors
such as $\eta^{\mu\nu}$. For example, if $\phi$ is an ordinary real
scalar field whose Lorentz tensor definition is independent of $\psi_{(a)}^{\mu}$,
frame independent theories can have $\psi_{(a)}^{\mu}$ only in the
combinations such as 
\begin{equation}
\sum_{ab}\eta^{ab}\psi_{(a)}^{\mu}\partial_{\mu}\phi\psi_{(b)}^{\nu}\partial_{\nu}\phi=\eta^{\mu\nu}\partial_{\mu}\phi\partial_{\nu}\phi
\end{equation}
where the $\psi_{(a)}^{\mu}$ disappears.

However, suppose we define $(\psi^{-1}x)^{\mu}\equiv(\Lambda^{-1})_{\phantom{\beta}\lambda}^{\beta}x^{\lambda}$
where $\psi_{(a)}^{\mu}\equiv\Lambda_{\phantom{\mu}a}^{\mu}$ in accordance
with Eq.~(\ref{eq:definingproperty}). Let's redefine a vector field
\begin{equation}
V^{\mu}(x)=\sum_{a}\phi^{(a)}(\psi^{-1}x)\psi_{(a)}^{\mu}\label{eq:linear}
\end{equation}
in terms of Lorentz scalar fields $\phi^{(a)}$ where we clearly see
that $\phi^{(a)}$ definition does depend on the definition of $\psi_{(a)}^{\mu}$.
Now, as long as $\phi^{(a)}(x)$ is used in the action, we cannot
absorb $\psi_{(a)}^{\mu}$ into a coordinate redefinition such that
$\psi_{(a)}^{\mu}$ completely disappears and at the same time maintain
Lorentz covariance. For example, the Lorentz and frame independent
$\int d^{4}x\chi(x)\partial_{\mu}V^{\mu}(x)$ where $\chi(x)$ is
a Lorentz scalar becomes 
\begin{align}
\int d^{4}x\chi(x)\partial_{\mu}V^{\mu}(x) & =\int d^{4}x\sum_{a}\chi(x)\psi_{(a)}^{\mu}\partial_{\mu}\phi^{(a)}(\psi^{-1}x)\\
 & =\sum_{a}\int d^{4}y\chi(\sum_{b}\psi_{(b)}^{\lambda}y^{b})\frac{\partial}{\partial y^{a}}\phi^{(a)}(y^{c})\label{eq:final}
\end{align}
where we defined 
\begin{equation}
dx^{\mu}=\sum_{b}\psi_{(b)}^{\mu}dy^{b}\label{eq:unusualbar}
\end{equation}
and used $d^{4}x=d^{4}y$. Even if we were to remove $\psi_{(b)}^{\lambda}$
in $\chi(\sum_{b}\psi_{(b)}^{\lambda}y^{b})$ through a field redefinition
of 
\begin{equation}
X(y)\equiv\chi(\sum_{b}\psi_{(b)}^{\lambda}y^{b})\label{eq:xdef}
\end{equation}
what remains is not a manifestly Lorentz invariant expression since
$\phi^{(a)}(y)$ is defined to be a scalar under Lorentz transformations.
On the other hand, it is clear from Eqs.~(\ref{eq:linear}) and (\ref{eq:final})
that manifest covariance can be recovered if one first considers $\psi_{(a)}^{\mu}=\delta_{\phantom{\mu}a}^{\mu}$
and now treats $a$ index as a covariantly transforming index (i.e.
$a$ transforming under the Lorentz group as $\bar{V}^{a}=\Lambda_{\phantom{a}c}^{a}V^{c}$).

Hence, we see that in situations in which there exists a frame in
which frame-dependent indices such as $(a)$ are contracted in a manifestly
covariant manner, the theory has no preferred frame as one can rewrite
it in a manifestly covariant notation. We will call this \textbf{\emph{frame-covariance}}.
The reason why this occurs is because Eq.~(\ref{eq:defofglobalfield})
tells us 
\begin{equation}
\psi_{(a)}^{\beta}=\Delta_{\phantom{\mu}\lambda}^{\mu}e_{\mu}[1](e_{\beta}[1])\otimes e^{\lambda}[1](e_{a}[1])
\end{equation}
which are just components of a $(1,1)$ Lorentz tensor in this defining
frame, just as in usual covariant notation where $a$ now behaves
as an ordinary Lorentz tensor index.

Next, there is another source of non-covariance besides the frame
choice. Although the functional degrees of freedom count between $\theta_{a}$
and $U_{a}$ is the same at least locally, the variable $\theta_{a}$
is different from $U_{a}$ from a frame-covariance representational
perspective because it has a non-linear map to ordinary tensor indices
unlike the index ``$a$'' on $U_{a}$. This non-linearity was introduced
to satisfy the gauge group representational requirement of the $G^{\mu\nu}$
while allowing an unconstrained path integration over $\theta_{a}$.
The loss of manifest covariance stems from introducing only the exact
number of functional degrees of freedom to match $A_{\mu}$. A more
precise representational theory explanation of these ``lost'' functional
degrees of freedom will be given in Sec.~\ref{sec:The-covariant-degrees}.

To better understand the nonlinearity that leads to the loss of frame-covariance
in terms of $\theta_{a}$, consider the analogous situation of 
\begin{equation}
V^{\mu}(x)=\sum_{a}\left[\Phi^{(a)}(\psi^{-1}x)\right]^{3}\psi_{(a)}^{\mu}\label{eq:nonlinear}
\end{equation}
where $\Phi^{(a)}(x)$ are scalar fields. Eq.~(\ref{eq:final}) turns
into 
\begin{equation}
\int d^{4}x\chi(x)\partial_{\mu}V^{\mu}(x)=\sum_{a}\int d^{4}yX(y)\frac{\partial}{\partial y^{a}}\left[\Phi^{(a)}(y)\right]^{3}
\end{equation}
where $X$ is defined in Eq.~(\ref{eq:xdef}). Unlike in Eq.~(\ref{eq:final}),
even if $a$ transforms as a Lorentz vector index starting from the
$\Lambda=1$ frame, one cannot recover manifest covariance because
of the nonlinearity of the index appearing in Eq.~(\ref{eq:nonlinear}).
Furthermore, the loss of frame-covariance also shows up in the path
integral measure: 
\begin{align}
DV= & D\Phi\det\frac{\delta V^{\mu}(x)}{\delta\Phi^{(c)}(y)}\\
= & D\Phi\det_{xy}\det_{\mu a}3\left[\Phi^{(a)}(\psi^{-1}x)\right]^{2}\psi_{(a)}^{\mu}\delta^{(4)}(x-y)\\
= & \mathcal{N}_{1}D\Phi\exp\left[i\int d^{4}yM^{4}\ln\left(\det_{\mu a}3\left[\Phi^{(a)}(y)\right]^{2}\psi_{(a)}^{\mu}\right)\right]
\end{align}
where $\mathcal{N}_{1}$ is a field independent normalization, $M^{4}$
is a momentum space regulator, and we used the change of variables
$y=\psi^{-1}x$. Hence, we see that 
\begin{eqnarray}
\int D\chi DVe^{i\int d^{4}x\chi(x)\partial_{\mu}V^{\mu}(x)} & = & \mathcal{N}\int DXD\Phi\exp\left[i\int d^{4}yM^{4}\ln\left(\det_{\mu a}3\left[\Phi^{(a)}(y)\right]^{2}\psi_{(a)}^{\mu}\right)\right.\nonumber \\
 &  & \left.+i\sum_{a}\int d^{4}yX(y)\frac{\partial}{\partial y^{a}}\left[\Phi^{(a)}(y)\right]^{3}\right]\label{eq:nonlinearexample}
\end{eqnarray}
where the right hand side not manifestly covariant and $\mathcal{N}$
is an unimportant normalization. Nonetheless the left hand side is
manifestly covariant and frame independent. In contrast, if we had
used the frame-covariant field redefinition Eq.~(\ref{eq:linear}),
we would have found 
\begin{equation}
\int D\chi DVe^{i\int d^{4}x\chi(x)\partial_{\mu}V^{\mu}(x)}\propto\int DXD\Phi\exp\left[i\int d^{4}yM^{4}\ln\left(\det_{\mu a}\psi_{(a)}^{\mu}\right)+i\sum_{a}\int d^{4}yX(y)\frac{\partial}{\partial y^{a}}\Phi^{a}(y)\right]\label{eq:framecovariant}
\end{equation}
which is manifestly frame-covariant (i.e.~manifestly Lorentz invariant
if both $a$ and $\mu$ indices are transformed under Lorentz transformations).
In this sense, the nonlinearity in the usage of the frame dependent
variable causes greater ``loss'' of manifest covariance. Note the
first term in the square bracket of Eq.~(\ref{eq:framecovariant})
(which can be dropped in computations since it is a field independent
term) is frame-covariant. That means in the frame-covariant field
redefinitions, we can recover the manifest Lorentz invariance as long
one moves to a Lorentz frame of $\psi_{(a)}^{\mu}=\delta_{\phantom{\mu}a}^{\mu}$
to eliminate $\psi$ and afterwards $a$ indices are transformed as
Lorentz tensor indices in Lorentz transformations.

A similar loss of manifest covariance occurs in BTGT theory \textbf{if
}one eliminates $\psi_{(a)}^{\mu}$ by a coordinate transformation
and uses $\theta_{a}^{B}$ fields which carry the frame index $a$
treated as a scalar field label. On the other hand, because the theory
is defined through a change of variables starting from the path integral
measure $DA_{\mu}(\theta_{a})$, the theory is covariant and frame
independent. This change of variables was mainly motivated by the
desire of rewriting the theory in terms of vierbeins and quantizing
it through an unconstrained path integration over $\theta_{a}$. Because
of the nonlinear field redefinition discussed in Eq.~(\ref{eq:nonlinearexample}),
frame-covariance does not really play a role in this argument (the
field redefinition argument starting from $A_{\mu}$) for frame independence
of the BTGT theory. On the other hand in Sec.~\ref{sec:Frame-independence-without}
we discuss how frame covariance can enter the frame independence of
a theory if we replace the crutch of using the field redefinition
$A_{\mu}(\theta_{a})$ with the symmetries of gauge invariance, BTGT
invariance, $S_{4}$ symmetry, and a particular parity symmetry of
the frame field.

\section{\label{sec:The-covariant-degrees}An extension of the BTGT symmetry}

The central object for BTGT in Eqs.~(\ref{eq:agrelation-1}) and
(\ref{eq:Gmunudef}) is the covariant tensor field $\left[G^{\mu\nu}(x)\right]^{ij}$.
In this section, we explain which manifestly covariant functional
degrees of freedom are thrown out when the BTGT theory is quantized
in terms of $\theta_{a}^{A}$ and the consequences of this in comparing
computations in different frames. In this way, we are able to identify
an extension to the previously stated BTGT symmetry, the equivalence
class of $\theta_{a}^{A}$ that give the same $A_{\mu}$. In particular,
the extension is to include the invariance with respect to different
frame choices $\{\psi_{(a)}^{\mu},x[1]\}$ and was already implicit
in previous papers \cite{Chung:2016lhv,Chung:2017zck,Basso:2019yap}.

We begin by giving a discussion of the Lorentz group representation
properties of $\left[G^{\mu\nu}(x)\right]^{ij}$. With the aim of
eventually accommodating Eqs.~(\ref{eq:Gtransform-1}) and (\ref{eq:agrelation-1}),
start with a general set of complex numbers $\left[G^{\mu\nu}\right]^{ij}$
where the $\mu,\nu$ indices transform through the usual real matrices
of the fundamental representation of $SO(1,3)$ and $i,j\in\{1,...,N\}$.\footnote{Here $N$ will eventually be identified with the gauge group generating
matrix $(T^{A})^{ij}$ being an $N\times N$ matrix as in Eq.~(\ref{eq:Gtransform-1}).} Because $[G^{\mu\nu}]^{ij}$ is required to be complex, for any fixed
indices $i,j$, $\left[G^{\mu\nu}\right]^{ij}$ is a $16\oplus16$
representation of $SO(1,3)$ where the second 16 of the direct sum
corresponds to the imaginary part of $\left[G^{\mu\nu}\right]^{ij}$.
The object $[G^{\mu\nu}]^{ij}$ has $(2\times16)N^{2}$ number of
real degrees of freedom.

In order to accommodate the choice of the exponentiation map Eq.~(\ref{eq:exponentialmap})
(motivated from the group representation property Eq.~(\ref{eq:Gtransform-1})),
we impose the constraint 
\begin{equation}
(G^{\alpha\nu})^{ji*}(G_{\alpha\mu})^{jk}=\delta_{\phantom{\nu}\mu}^{\nu}\delta^{ik}\label{eq:u13induced}
\end{equation}
where the Latin indices such as $i,j$ refer to the space spanned
by $i,j$ indices in the generator matrix $T_{ij}^{B}$ with $i,j\in\{1,...,N\}$.
One can easily show that this constraint corresponds to restricting
$(G_{\phantom{\lambda}\mu}^{\lambda})^{jk}$ to having $16N^{2}$
real degrees of freedom . Hence, imposing this constraint reduced
the number of degrees of freedom from $32N^{2}$ to $16N^{2}$. Indeed
$(G_{\phantom{\alpha}\mu}^{\alpha})^{jk}$ can be thought of elements
of $U(N,3N)$. Since Eq.~(\ref{eq:u13induced}) is a covariant constraint,
manifest Lorentz invariance has been preserved. 

Since we want the smallest $(G_{\phantom{\alpha}\mu}^{\alpha})^{jk}$
representation to identify the smallest set of manifestly covariant
degrees of freedom being thrown out when the BTGT theory is quantized
in terms of $\theta_{a}$, we can then ask whether we can reduce the
$16N^{2}$ degrees of freedom further closer to $4N^{2}$ without
sacrificing covariance. To this end, note the raised Lorentz index
matrix $(G^{\alpha\beta})^{ij}$ in the ansatz Eq.~(\ref{eq:exponentialmap})
is symmetric in the Lorentz indices. We also know that any Lorentz
transform orbits of $(G^{\alpha\beta})^{ij}$ that is symmetric in
the Lorentz indices remains symmetric. However, we cannot simply conclude
from the usual $SO(1,3)$ decomposition $16=1\oplus3\oplus3\oplus9$
with the $10=1\oplus9$ being symmetric that a possible representation
is $10N^{2}\oplus10N^{2}$ since we need to satisfy Eq.~(\ref{eq:u13induced}).
Indeed, we already know the restriction gives a smaller set since
$16N^{2}<20N^{2}$. We can therefore try to solve Eq.~(\ref{eq:u13induced})
explicitly and impose the restriction of index symmetry.

Start by noting that a $U(1,3)$ group representation matrices are
the set of matrices $U$ satisfying 
\begin{equation}
U^{\dagger}\eta U=\eta
\end{equation}
or equivalently
\begin{equation}
U^{\alpha\nu*}U_{\alpha\mu}=\delta_{\phantom{\nu}\mu}^{\nu}
\end{equation}
where $\eta$ is the Lorentz metric matrix reminiscent of Eq.~(\ref{eq:u13induced}).
Hence, we will try to find $(G^{\alpha\beta})^{ij}$ satisfying exchange
symmetry in the Greek indices and Eq.~(\ref{eq:u13induced}) by starting
with the $U(1,3)$ group matrix fundamental representation and then
making each group matrix element be valued in a gauge Lie algebra
matrix space in an appropriate way. As we will see below, these conditions
will reduce the number of degrees of freedom from $16N^{2}$ to $10D(\mathcal{G})$
where $D(\mathcal{G})\leq N^{2}$ is the dimension of the gauge group
transforming according to Eq.~(\ref{eq:Gtransform-1}).

The fundamental representation of $U(1,3)$ can be explicitly generated
by the 16 matrices (elements of the Lie algebra $u(1,3)$) 
\begin{eqnarray}
S\equiv\{\Xi^{1},\Xi^{2},...,\Xi^{10}\} & \equiv & \left\{ \left(\begin{array}{cccc}
1 & 0 & 0 & 0\\
0 & 0 & 0 & 0\\
0 & 0 & 0 & 0\\
0 & 0 & 0 & 0
\end{array}\right),\left(\begin{array}{cccc}
0 & 0 & 0 & 0\\
0 & -1 & 0 & 0\\
0 & 0 & 0 & 0\\
0 & 0 & 0 & 0
\end{array}\right),\left(\begin{array}{cccc}
0 & 0 & 0 & 0\\
0 & 0 & 0 & 0\\
0 & 0 & -1 & 0\\
0 & 0 & 0 & 0
\end{array}\right),\left(\begin{array}{cccc}
0 & 0 & 0 & 0\\
0 & 0 & 0 & 0\\
0 & 0 & 0 & 0\\
0 & 0 & 0 & -1
\end{array}\right),\right.\nonumber \\
 &  & \left(\begin{array}{cccc}
0 & -1 & 0 & 0\\
1 & 0 & 0 & 0\\
0 & 0 & 0 & 0\\
0 & 0 & 0 & 0
\end{array}\right),\left(\begin{array}{cccc}
0 & 0 & -1 & 0\\
0 & 0 & 0 & 0\\
1 & 0 & 0 & 0\\
0 & 0 & 0 & 0
\end{array}\right),\left(\begin{array}{cccc}
0 & 0 & 0 & -1\\
0 & 0 & 0 & 0\\
0 & 0 & 0 & 0\\
1 & 0 & 0 & 0
\end{array}\right),\left(\begin{array}{cccc}
0 & 0 & 0 & 0\\
0 & 0 & 1 & 0\\
0 & 1 & 0 & 0\\
0 & 0 & 0 & 0
\end{array}\right),\nonumber \\
 &  & \left.\left(\begin{array}{cccc}
0 & 0 & 0 & 0\\
0 & 0 & 0 & 1\\
0 & 0 & 0 & 0\\
0 & 1 & 0 & 0
\end{array}\right),\left(\begin{array}{cccc}
0 & 0 & 0 & 0\\
0 & 0 & 0 & 0\\
0 & 0 & 0 & 1\\
0 & 0 & 1 & 0
\end{array}\right)\right\} 
\end{eqnarray}
and 
\begin{eqnarray}
J\equiv\{\Xi^{11},\Xi^{12},...,\Xi^{16}\} & \equiv & \left\{ i\left(\begin{array}{cccc}
0 & 1 & 0 & 0\\
1 & 0 & 0 & 0\\
0 & 0 & 0 & 0\\
0 & 0 & 0 & 0
\end{array}\right),i\left(\begin{array}{cccc}
0 & 0 & 1 & 0\\
0 & 0 & 0 & 0\\
1 & 0 & 0 & 0\\
0 & 0 & 0 & 0
\end{array}\right),i\left(\begin{array}{cccc}
0 & 0 & 0 & 1\\
0 & 0 & 0 & 0\\
0 & 0 & 0 & 0\\
1 & 0 & 0 & 0
\end{array}\right),\right.\nonumber \\
 &  & \left.i\left(\begin{array}{cccc}
0 & 0 & 0 & 0\\
0 & 0 & -1 & 0\\
0 & 1 & 0 & 0\\
0 & 0 & 0 & 0
\end{array}\right),i\left(\begin{array}{cccc}
0 & 0 & 0 & 0\\
0 & 0 & 0 & 1\\
0 & 0 & 0 & 0\\
0 & -1 & 0 & 0
\end{array}\right),\,i\left(\begin{array}{cccc}
0 & 0 & 0 & 0\\
0 & 0 & 0 & 0\\
0 & 0 & 0 & 1\\
0 & 0 & -1 & 0
\end{array}\right)\right\} 
\end{eqnarray}
where the matrix entries correspond to Lorentz $\mu\nu$ entries.
Although the matrices generated by $J$ form the group $SO(1,3)$
subgroup while the matrices generated by $S$ do not form a group,
this does not mean we should restrict to the subgroup generated by
$J$ for constructing BTGT because we are not looking for the group
matrices but the representation basis tensors.\footnote{An $SU(2)$ group matrix can be represented as $\Gamma^{ij}\equiv\exp(i\theta^{A}\sigma^{A}/2)^{ij}$
while the the basis tensor for this representation is $\psi^{i}$
with the orbit $\psi'=\Gamma\psi$.} Now, we impose the symmetry of $(G^{\alpha\beta})^{ij}=(G^{\beta\alpha})^{ij}$
(imposed by consistency with the ansatz Eq.~(\ref{eq:exponentialmap}))
as planned by choosing $S$ to generate $(G_{\phantom{\alpha}\beta}^{\alpha})^{ij}$.
First, note
\begin{equation}
K_{\phantom{\alpha}\beta}^{\alpha}=\exp\left[i\sum_{a=1}^{10}\epsilon_{a}(x)\Xi^{a}\right]_{\phantom{\alpha}\beta}^{\alpha}
\end{equation}
gives a symmetric $K^{\alpha\beta}$. This property can be imported
to describing $(G^{\alpha\beta})^{jk}$. To solve Eq.~(\ref{eq:u13induced}),
we still have to impose the condition on the Latin indices of $(G_{\phantom{\alpha}\mu}^{\alpha})^{jk}$.
To see if the solution is possible, try
\begin{equation}
(G_{\phantom{\alpha}\beta}^{\alpha})^{ij}=\exp\left[i\sum_{a=1}^{10}\alpha_{a}(x)\Xi^{a}\right]_{\phantom{\alpha}\beta}^{\alpha}\label{eq:10exp}
\end{equation}
where $\alpha_{a}$ is Lie algebra valued: i.~e.~ 
\begin{equation}
\alpha_{a}\equiv\sum_{C}\alpha_{a}^{C}T^{C}
\end{equation}
and the product $T^{C}\Xi^{a}T^{E}\Xi^{b}$ is defined as in the usual
tensor product space: 
\begin{equation}
\left[\left(T^{C}\Xi^{a}T^{E}\Xi^{b}\right)_{\phantom{\alpha}\beta}^{\alpha}\right]^{ij}=\left(T^{C}\right)^{ik}\left(T^{E}\right)^{kj}\left(\Xi^{a}\right)_{\phantom{\alpha}\mu}^{\alpha}\left(\Xi^{b}\right)_{\phantom{\mu}\beta}^{\mu}.
\end{equation}
It is easy to check that Eq.~(\ref{eq:u13induced}) can be satisfied
as follows. First, expand: 
\begin{align}
(G^{\alpha\nu})^{ji*}(G_{\alpha\mu})^{jk} & =\sum_{\omega\alpha\lambda j}(G_{\phantom{\alpha}\omega}^{\alpha}\eta^{\omega\nu})^{ji*}\eta_{\alpha\lambda}(G_{\phantom{\lambda}\mu}^{\lambda})^{jk}\\
 & =\sum_{j}\eta(\exp\left[-i\sum_{a=1}^{10}\alpha_{a}(x)\Xi^{a\dagger}\right])^{ij}\eta(\exp\left[i\sum_{a=1}^{10}\alpha_{a}(x)\Xi^{a}\right])^{jk}.
\end{align}
Using the property $\eta\Xi^{a\dagger}\eta=\Xi^{a}$, we conclude
this is equivalent to Eq.~(\ref{eq:u13induced}). 

Even though $T^{C}\Xi^{a}$ is not in general a group generator, we
are most concerned with whether the orbit of $(G_{\phantom{\alpha}\beta}^{\alpha})^{ij}$
under Lorentz transformations and can be written in the form of Eq.~(\ref{eq:10exp}).
Note that there generically exists a solution $\tilde{\alpha}_{a}^{A}(x)$
to 
\begin{equation}
\left(\exp\left[i\sum_{a=1}^{10}\sum_{A}\alpha_{a}^{A}(x)T^{A}\Xi^{a}\right]_{\phantom{\alpha}\beta}^{\alpha}\right)^{ij}\exp\left[i\sum_{B}\Theta^{B}(x)T^{B}\right]^{jk}=\left(\exp\left[i\sum_{a=1}^{10}\sum_{A}\tilde{\alpha}_{a}^{A}(x)T^{A}\Xi^{a}\right]_{\phantom{\alpha}\beta}^{\alpha}\right)^{ik}
\end{equation}
since 
\begin{equation}
[T^{A}\Xi^{a},T^{B}\mathbb{I}]=if^{ABC}T^{C}\Xi^{a}.
\end{equation}
Furthermore, since
\begin{equation}
\exp\left[i\sum_{a=1}^{10}\sum_{A}\alpha_{a}^{A}(x)T^{A}\Xi^{a}\right]^{\alpha\beta}=\exp\left[i\sum_{a=1}^{10}\sum_{A}\alpha_{a}^{A}(x)T^{A}\Xi^{a}\right]^{\beta\alpha}
\end{equation}
(with the $T^{A}$ associated Latin indices suppressed) and the Lorentz
transform induced orbits preserve this symmetry, one generically expects
there to be a solution to
\begin{equation}
\Lambda_{\phantom{\mu}\alpha}^{\mu}\Lambda_{\phantom{\mu}\beta}^{\nu}\exp\left[i\sum_{a=1}^{10}\sum_{A}\alpha_{a}^{A}(x)T^{A}\Xi^{a}\right]^{\alpha\beta}=\exp\left[i\sum_{a=1}^{10}\sum_{B}\bar{\alpha}_{a}^{B}(\Lambda x)T^{B}\Xi^{a}\right]^{\mu\nu}
\end{equation}
as we will now verify. 

For verification, we give explicit representation of the covariant
$10\times D(A)$ degrees of freedom in $\alpha_{a}^{A}(x)T^{A}\Xi^{a}$,
which can be given in many forms. One form is 
\begin{equation}
K_{\phantom{\mu}\nu}^{\mu}(x)=\exp\left[i\sum_{a=1}^{10}\alpha_{a}(x)\Xi^{a}\right]
\end{equation}
which under Lorentz transformation transforms as
\begin{equation}
\bar{\alpha}_{c}(\bar{x})=\frac{1}{N_{c}}\sum_{a}\alpha_{a}(\Lambda^{-1}\bar{x}){\rm Tr}\left(\Xi^{c}\Lambda\Xi^{a}\Lambda^{-1}\right)
\end{equation}
\begin{equation}
{\rm Tr}\left(\Xi^{c}\Xi^{f}\right)=\delta^{cf}N_{c}
\end{equation}
where $\alpha_{a}$ is Lie algebra valued: i.~e.~ 
\begin{equation}
\alpha_{a}\equiv\sum_{C}\alpha_{a}^{C}T^{C}
\end{equation}
where if $\xi_{(f)}^{*l}\left(K_{\phantom{\mu}\nu}^{\mu}\right)^{lm}$
is to transform as an antifundamental, we can set $T^{C}$ equal to
the negative of the Hermitian conjugate of the fundamental representation
matrices and sum over $D(A)$ of them. Note also that the trace here
is only over the Lorentz spacetime indices. This form is useful in
counting the degrees of freedom and its concise relationship to the
generators of $u(1,3)$. A more familiar Lorentz tensorial form is
given by making a variable change $\alpha_{a}\rightarrow\alpha^{\alpha\beta}=\alpha^{(\alpha\beta)}$
where 
\begin{equation}
\alpha_{a}(x)\left(\Xi^{a}\right)^{\mu\nu}=\frac{1}{2}\alpha^{\alpha\beta}\left(\Xi_{\alpha\beta}\right)_{\phantom{\mu}\lambda}^{\mu}\eta^{\lambda\nu}
\end{equation}
\begin{equation}
\alpha_{a}(x)\left(\Xi^{a}\right)_{\phantom{\mu}\lambda}^{\mu}\eta^{\lambda\nu}=\alpha^{\alpha\beta}\delta_{\phantom{\mu}(\alpha}^{\mu}\delta_{\phantom{\nu}\beta)}^{\nu}
\end{equation}
and 
\begin{eqnarray}
\{\Xi^{00}\eta,\Xi^{11}\eta,\Xi^{22}\eta,\Xi^{33}\eta\} & \equiv & \left\{ 2\left(\begin{array}{cccc}
1 & 0 & 0 & 0\\
0 & 0 & 0 & 0\\
0 & 0 & 0 & 0\\
0 & 0 & 0 & 0
\end{array}\right),2\left(\begin{array}{cccc}
0 & 0 & 0 & 0\\
0 & -1 & 0 & 0\\
0 & 0 & 0 & 0\\
0 & 0 & 0 & 0
\end{array}\right),2\left(\begin{array}{cccc}
0 & 0 & 0 & 0\\
0 & 0 & 0 & 0\\
0 & 0 & -1 & 0\\
0 & 0 & 0 & 0
\end{array}\right),\right.\nonumber \\
 &  & \left.2\left(\begin{array}{cccc}
0 & 0 & 0 & 0\\
0 & 0 & 0 & 0\\
0 & 0 & 0 & 0\\
0 & 0 & 0 & -1
\end{array}\right)\right\} \\
\{\Xi^{01}\eta,\Xi^{10}\eta,\Xi^{02}\eta,\Xi^{20}\eta,..\} & \equiv & \left\{ \left(\begin{array}{cccc}
0 & 1 & 0 & 0\\
0 & 0 & 0 & 0\\
0 & 0 & 0 & 0\\
0 & 0 & 0 & 0
\end{array}\right),\left(\begin{array}{cccc}
0 & 0 & 0 & 0\\
1 & 0 & 0 & 0\\
0 & 0 & 0 & 0\\
0 & 0 & 0 & 0
\end{array}\right),\left(\begin{array}{cccc}
0 & 0 & 1 & 0\\
0 & 0 & 0 & 0\\
0 & 0 & 0 & 0\\
0 & 0 & 0 & 0
\end{array}\right),\right.\nonumber \\
 &  & \left.\left(\begin{array}{cccc}
0 & 0 & 0 & 0\\
0 & 0 & 0 & 0\\
1 & 0 & 0 & 0\\
0 & 0 & 0 & 0
\end{array}\right),...\right\} .
\end{eqnarray}
With this definition, the fields $\alpha^{\mu\nu}$ have the following
Lorentz transformation property: 
\begin{equation}
\alpha^{\kappa\phi}(\Lambda^{-1}\bar{x})\Lambda_{\phantom{\mu}\kappa}^{\mu}\Lambda_{\phantom{\delta}\phi}^{\nu}=\bar{\alpha}^{\mu\nu}(\bar{x})
\end{equation}
which is obviously useful for constructing field theories. From this,
it is clear that one can recover a manifestly frame independent description
with $10\times D(\mathcal{G})$ degrees of freedom whereas the diagonal
fields $\{\alpha^{00},\alpha^{11},\alpha^{22},\alpha^{33}\}$ are
the $\theta_{a}$ degrees of freedom.

Since $S$ has 10 elements, we see that we can describe a manifestly
frame independent covariant complex tensor $(G_{\phantom{\alpha}\mu}^{\alpha})^{jk}$
with $10\times D(\mathcal{G})$ real degrees of freedom. The subset
of $S$ given by $\{\Xi_{1},\Xi_{2},\Xi_{3},\Xi_{4}\}$ (the Cartan
subalgebra of $u(1,3)$) generate the Abelian group $U(1)^{4}$ as
usual, and it is this group manifold that is parameterized by the
BTGT fields $\theta_{a}^{B}$ for any fixed $B$. Hence, we have arrived
at our answer: it is the 6$D(\mathcal{G})$ related to the non-Cartan-subalgebra
elements of $u(1,3)$ in $S$ that are lost in using the $\theta_{a}^{B}$
ansatz Eq.~(\ref{eq:exponentialmap}) and leads to a frame-dependent
description of a frame independent theory. 

One consequence of describing the covariant frame independent theory
using only $4\times D(\mathcal{G})$ degrees of freedom (instead of
$10\times D(\mathcal{G})$) is that when one compares $G^{\mu\nu}$
computations executed in two different $x[1]$ frame choices (see
Eq.~(\ref{eq:defofbracket}) for the definition of $x[1]$), one
is actually considering different tensors and not the components of
the same tensor in different Lorentz frames. To be more precise, suppose
one $x[1]$ choice (with coordinate basis $e_{\nu}$) is related to
the other $\underline{x}[1]$ choice through a Lorentz transform $\underline{x}[1]=x[2]=\Lambda x[1]$.
Suppose the field $G^{\mu\nu}$ in $x[1]$ coordinates is computed
with a basis choice $\psi_{(a)}^{\mu}$: 
\begin{equation}
G_{\phantom{\mu}\nu}^{\mu}(x[1])e_{\mu}\otimes e^{\nu}=\left(\exp[-i\sum_{a}\theta_{a}(x[1])\psi_{(a)}\psi_{(a)}\eta^{aa}]\right)_{\phantom{\mu}\nu}^{\mu}e_{\mu}\otimes e^{\nu}
\end{equation}
In $x[2]=\Lambda x[1]$ coordinates where $\bar{\psi}_{(a)}^{\mu}=\Lambda_{\phantom{\mu}\lambda}^{\mu}\psi_{(a)}^{\lambda}$
and $\bar{e}_{\mu}=(\Lambda^{-1})_{\phantom{\beta}\mu}^{\beta}e_{\beta}$,
we know 
\begin{equation}
G_{\phantom{\mu}\nu}^{\mu}(x[1])e_{\mu}\otimes e^{\nu}=\bar{G}_{\phantom{\mu}\nu}^{\mu}(x[2])\bar{e}_{\mu}\otimes\bar{e}^{\nu}=\left(\exp[-i\sum_{a}\theta_{a}(\Lambda^{-1}x[2])\bar{\psi}_{(a)}\bar{\psi}_{(a)}\eta^{aa}]\right)_{\phantom{\mu}\nu}^{\mu}\bar{e}_{\mu}\otimes\bar{e}^{\nu}
\end{equation}
where we have used the scalar transformation property of $\theta_{a}$.
Now, suppose one considers the same choice of basis $\psi_{(a)}^{\mu}$
in $\underline{x}[1]=x[2]$ frame and executes the computation. One
can easily obtain 
\begin{equation}
\underline{G}_{\phantom{\mu}\nu}^{\mu}(x[2])\bar{e}_{\mu}\otimes\bar{e}^{\nu}=\left(\exp[-i\sum_{a}\Theta_{a}(x[2])\psi_{(a)}\psi_{(a)}\eta^{aa}]\right)_{\phantom{\mu}\nu}^{\mu}\bar{e}_{\mu}\otimes\bar{e}^{\nu}
\end{equation}
where 
\begin{equation}
\sum_{a}\Theta_{a}(x[2])\psi_{(a)}^{\alpha}\psi_{(a)}^{\beta}\eta^{aa}\neq\sum_{a}\theta_{a}(\Lambda^{-1}x[2])\bar{\psi}_{(a)}^{\alpha}\bar{\psi}_{(a)}^{\beta}\eta^{aa}.
\end{equation}
This means 
\begin{equation}
\underline{G}_{\phantom{\mu}\nu}^{\mu}(x[2])\bar{e}_{\mu}\otimes\bar{e}^{\nu}\neq\bar{G}_{\phantom{\mu}\nu}^{\mu}(x[2])\bar{e}_{\mu}\otimes\bar{e}^{\nu}\label{eq:equiv1}
\end{equation}
even though 
\begin{equation}
\left[\underline{G}^{-1\alpha\beta}(x[2])\right]\left[\frac{\partial}{\partial x^{\alpha}[2]}\underline{G}_{\beta\mu}(x[2])\right]\bar{e}^{\mu}=\left[\bar{G}^{-1\alpha\beta}(x[2])\right]\left[\frac{\partial}{\partial x^{\alpha}[2]}\bar{G}_{\beta\mu}(x[2])\right]\bar{e}^{\mu}.\label{eq:equiv2}
\end{equation}
Consequently, the map of $G_{\phantom{\alpha}\beta}^{\alpha}$ to
$A_{\mu}$ together with the covariance of $G_{\phantom{\alpha}\beta}^{\alpha}$
and $A_{\mu}$ induce an equivalence class of of tensors $G^{\mu\nu}e_{\mu}\otimes e_{\nu}\sim\underline{G}^{\alpha\beta}\bar{e}_{\alpha}\otimes\bar{e}_{\beta}$
even though $G^{\mu\nu}e_{\mu}\otimes e_{\nu}=\bar{G}^{\mu\nu}\bar{e}_{\mu}\otimes\bar{e}_{\nu}\neq\underline{G}^{\alpha\beta}\bar{e}_{\alpha}\otimes\bar{e}_{\beta}$.

Although we already know that BTGT transform is identified as an $A_{\mu}$
equivalence class, what we learn from this section is that the equivalence
class is a combination of an $\{\psi_{(a)}^{\mu},x[1]\}$ choice \emph{in
addition to} the BTGT transform. That is because the BTGT symmetry
transformations (see Eq.~(\ref{eq:BTGT sym for Ua})) defining a
set of equivalence class conditions involve fewer than $4\times D(\mathcal{G})$
real functional degrees of freedom since according to the constraint
Eq.~(\ref{eq:Zconstraint}), the equivalence class functions in the
BTGT are functions of one fewer dimension (i.e. 3-dimensional functions
in a 4-dimensional spacetime) which is certainly fewer than the $6\times D(\mathcal{G})$
degrees of freedom required to restore manifest frame independence.
The data of $\{\psi_{(a)}^{\mu},x[1]\}$ chooses which $4\times D(\mathcal{G})$
of the $10\times D(\mathcal{G})$ real functional degrees of freedom
that one is calling $\theta_{a}^{B}$. BTGT is automatically invariant
under $\{\psi_{(a)}^{\mu},x[1]\}$ choice since the path integral
is defined originally with respect to $A_{\mu}$. 

One might wonder whether working with $\theta_{a}$ frame dependent
fields will require a strong frame dependence of the counter terms.
In our previous explicit computations \cite{Chung:2017zck,Basso:2019yap},
we saw no evidence for the variable change playing a role at one-loop.
As we will see in the next section, there is a good reason why this
change of variables does not lead to disastrous loss of frame invariance
through the counter terms: the combination of BTGT, gauge, and Lorentz
invariance, together with $S_{4}$ symmetry and a parity symmetry
of the frame field\lyxdeleted{danielchung}{Mon Dec 30 08:25:17 2019}{
} implicit in the BTGT formulation leads to manifest frame independence
of the action.

\section{\label{sec:Frame-independence-without}Frame independence without
relying on the $A_{\mu}$ field}

Thus far, we have focused on the frame independence of BTGT being
a consequence of a field redefinition of the manifestly covariant
field $A_{\mu}(\theta_{a})$. We know that $A_{\mu}$ is required
for gauge invariance. Furthermore, the relationship between $\theta_{a}$
and $A_{\mu}$ is constrained by BTGT invariance, gauge invariance,
and the choice of $\{\psi_{(a)}^{\mu},x[1]\}$. In this section, we
show how to remove the crutch of $A_{\mu}$ field (at the action level)
with two additional conditions to gauge and BTGT invariance. More
explicitly, we show how to construct theories with $\{\psi_{(a)}^{\mu},x[1]\}$
choices that do not have any physical effect on the theory (except
possibly for zero modes) if we add the following conditions to the
theory: an $S_{4}$ permutation symmetry generated by exchanges 
\begin{equation}
\left\{ \eta^{aa},\psi_{(a)},U_{a}\right\} \leftrightarrow\left\{ \eta^{bb},\psi_{(b)},U_{b}\right\} \label{eq:perm sym frame ind proof}
\end{equation}
 for any BTGT indices $a$ and $b$ in $\{0,...,3\}$, and a rigid
rescaling symmetry
\begin{equation}
\psi_{(a)}^{\mu}\rightarrow\lambda_{a}\psi_{(a)}^{\mu}\label{eq:rescaling sym frame ind proof}
\end{equation}
of the basis fields for any constant $\lambda_{a}\neq0$. This construction
is done without starting from the theory constructed from gauge connection
$A_{\mu}$.

First, we begin with some comments about conventions chosen. Via the
rescaling symmetry of Eq.~(\ref{eq:rescaling sym frame ind proof}),
one can normalize the orthogonal basis set $\psi_{(a)}^{\mu}$ such
that 
\begin{equation}
(\psi_{(a)}^{\lambda}\psi_{(a)\lambda})^{-1}=\eta^{aa}\label{eq:psi normalized is eta}
\end{equation}
without loss of generality. After normalizing, there remains a residual
$\psi_{(a)}^{\mu}$ parity symmetry 
\begin{equation}
\psi_{(a)}^{\mu}\rightarrow\mbox{sign}\left(\lambda_{a}\right)\psi_{(a)}^{\mu}\label{eq:psi parity PT discrete sym}
\end{equation}
that corresponds to the $PT$ discrete subgroup of the $O(1,3)$ Lorentz
group. The choice of Eq.~(\ref{eq:psi normalized is eta}) is implied
for the rest of this section. In addition, although $\theta_{a}$
and $U_{a}=e^{i\theta_{a}}$ are functionally equivalent, the advantage
of $U_{a}$ is that the symmetry transformations are tensorial (by
construction of BTGT), while the non-Abelian $\theta_{a}$ transforms
inhomogeneously and nonlinearly. Constructing BTGT and gauge invariants
is therefore simpler when working with $U_{a}$.

The proposition we would like to establish in this section is the
following. Suppose the BTGT and gauge invariant action $S\left[\psi_{(a)}^{\mu},U_{a},\phi\right]$
is invariant under the symmetries of Eqs.~(\ref{eq:perm sym frame ind proof})
and (\ref{eq:rescaling sym frame ind proof}), where $\phi$ is a
scalar matter field transforming under the fundamental representation
of gauge group $\mathcal{G}$. Let the action $S$ be derived from
a local Lagrangian $\mathcal{L}$ and renormalizable. Given these
conditions, the action is frame independent as defined in Appendix
\ref{sec:Definitions}. Explicitly, there exists a change of variables
$V^{\mu}\left(\psi_{(a)}^{\mu},U_{a}\right)$ and a new action $\tilde{S}[V^{\mu},\phi]$
that is independent of $\psi_{(a)}^{\mu}$ such that 
\begin{equation}
S\left[\psi_{(a)}^{\mu},U_{a},\phi\right]=\tilde{S}\left[V^{\mu},\phi\right].\label{eq:btgt action is vector gauge action}
\end{equation}

The intuition is that gauge invariance, BTGT invariance, and the conditions
of Eqs.~(\ref{eq:perm sym frame ind proof}) and (\ref{eq:rescaling sym frame ind proof})
combine to restrict the theory to only being constructed from gauge
covariant derivatives of the form 
\begin{equation}
D^{\mu}\left(\boldsymbol{\cdot}\right)=\sum_{a}\eta^{aa}\psi_{(a)}^{\mu}\psi_{(a)}^{\lambda}U_{a}\partial_{\lambda}\left(U_{a}^{-1}\boldsymbol{\cdot}\right)\label{eq:covariant derivative, c is eta}
\end{equation}
where the argument $\boldsymbol{\cdot}$ stands for any field in some
representation $R$ of the gauge group and $U_{a}=e^{i\theta_{a}^{A}T_{(R)}^{A}}$,
where $T_{(R)}^{A}$ are the basis elements of the Lie algebra in
representation $R$. Eq.~(\ref{eq:covariant derivative, c is eta})
has a decomposition into $\partial^{\mu}$ and vector field 
\begin{equation}
V^{\mu}(\psi_{(a)}^{\mu},U_{a})=i\sum_{a}\eta^{aa}\psi_{(a)}^{\mu}\psi_{(a)}^{\lambda}U_{a}\partial_{\lambda}\left(U_{a}^{-1}\right).\label{eq:V in terms of Ua}
\end{equation}
Due to gauge invariance, the Lagrangian must be composed from covariant
derivatives of Eq.~(\ref{eq:covariant derivative, c is eta}). The
only $U_{a}$ dependence of the action $S\left[\psi_{(a)}^{\mu},U_{a},\phi\right]$
is then of the form Eq.~(\ref{eq:V in terms of Ua}). Therefore we
end up with an action $\tilde{S}\left[\psi_{(a)}^{\mu},V^{\mu}(\psi_{(a)}^{\mu},U_{a}),\phi\right]$
that satisfies Eq.~(\ref{eq:btgt action is vector gauge action}).
The residual $\psi_{(a)}^{\mu}$ frame dependence of $\tilde{S}$
is finally removed by restricting to renormalizable terms consistent
with the parity and $S_{4}$ symmetry properties given in Eqs.~(\ref{eq:psi parity PT discrete sym})
and (\ref{eq:perm sym frame ind proof}).

Let us now proceed with the details. In Appendix \ref{sub:Lemma of Ua dependence},
it is shown that the most general BTGT invariant Lagrangian must be
a function of BTGT invariant monomials $U_{a}\partial_{(a)}U_{a}^{-1}$
and $\partial_{\mu}\left(U_{a}\partial_{(a)}U_{a}^{-1}\right)$. Explicitly,
the functional dependence of the Lagrangian is

\begin{equation}
\mathcal{L}=\mathcal{L}\left(\psi_{(a)}^{\mu},U_{a}\partial_{(a)}U_{a}^{-1},\partial_{\mu}\left(U_{a}\partial_{(a)}U_{a}^{-1}\right),\phi,\partial_{\mu}\phi,\phi^{\dagger},\partial_{\mu}\phi^{\dagger}\right),
\end{equation}
where $\partial_{(a)}U_{a}^{-1}\equiv\psi_{(a)}^{\mu}\partial_{\mu}U_{a}^{-1}$
is the projected derivative. We next use gauge invariance to further
restrict the theory. The gauge covariant derivative which has the
usual transformation properties is

\begin{equation}
D^{\mu}(\boldsymbol{\cdot})=\sum_{a,b}c^{ab}\psi_{(a)}^{\mu}U_{b}\partial_{(b)}\left(U_{b}^{-1}\boldsymbol{\cdot}\right),\label{eq:covariant derivative with c ab}
\end{equation}
where any choice of the constant $c^{ab}$ is compatible with both
gauge and BTGT invariance. Imposing the conditions of Eqs.~(\ref{eq:perm sym frame ind proof})
and (\ref{eq:psi parity PT discrete sym}) restricts this to 
\begin{equation}
c^{ab}=\delta^{ab}\eta^{aa}=\eta^{ab},\label{eq:c is eta condition}
\end{equation}
which corresponds to the covariant derivative of Eq.~(\ref{eq:covariant derivative, c is eta}).
Given Eq.~(\ref{eq:c is eta condition}), the projected covariant
derivative of $U_{a}$ is zero: $D_{(a)}U_{a}=\psi_{(a)\mu}D^{\mu}U_{a}=U_{a}\partial_{(a)}\left(U_{a}^{-1}U_{a}\right)=0.$
Any term containing such a derivative is therefore zero after imposing
gauge invariance by promoting ordinary derivatives to covariant derivatives
defined by Eq.~(\ref{eq:covariant derivative, c is eta}). Since
BTGT invariance requires projected derivatives $\partial_{(a)}U_{a}^{-1}$,
all terms involving projected derivatives of $U_{a}$ go to zero once
the theory is made gauge covariant. The terms consistent with gauge
invariance are then composed of $F^{\mu\nu}\equiv i\left[D^{\mu},D^{\nu}\right]$
, $D^{\mu}\phi$, and its conjugate $D^{\mu}\phi^{\dagger}$. Explicitly
terms of fields, 
\begin{align}
F^{\mu\nu} & =i\sum_{a,b}\eta^{aa}\eta^{bb}\psi_{(a)}^{[\mu}\psi_{(b)}^{\nu]}U_{a}\partial_{(a)}\left(U_{a}^{-1}U_{b}\partial_{(b)}\left(U_{b}^{-1}\right)\right)\label{eq:F btgt formalism}\\
 & =\partial^{[\mu}V^{\nu]}(\psi_{(a)}^{\mu},U_{a})-iV^{[\mu}(\psi_{(a)}^{\mu},U_{a})V^{\nu]}(\psi_{(a)}^{\mu},U_{a}),\label{eq:F in terms of vector}
\end{align}
 and 
\begin{align}
D^{\mu}\phi & =\sum_{a}\eta^{aa}\psi_{(a)}^{\mu}U_{a}\partial_{(a)}\left(U_{a}^{-1}\phi\right)\\
 & =\partial^{\mu}\phi-iV^{\mu}(\psi_{(a)}^{\mu},U_{a})\phi.\label{eq:Dphi in terms of vector}
\end{align}
where $V^{\mu}(\psi,U)$ is defined in Eq.~(\ref{eq:V in terms of Ua}).
It is interesting that the commutator of fields does not appear in
the BTGT formalism Eq.~(\ref{eq:F btgt formalism}). More precisely,
the gauge curvature tensor $F^{\mu\nu}$ is homogeneous in ordinary
derivatives, just like in the Abelian theory. We defer the exploration
of this property to a future work.

The $\psi_{(a)}^{\mu}$ parity symmetry of Eq.~(\ref{eq:psi parity PT discrete sym})
only allows $\psi_{(a)}^{\mu}$ to appear in the bilinear invariant
$(H^{a})^{\mu\nu}$ of Eq.~(\ref{eq:intermsofpsi}). The most general
BTGT and gauge invariant renormalizable Lagrangian consistent with
the $\psi_{(a)}^{\mu}$ parity symmetry of Eq.~(\ref{eq:psi parity PT discrete sym})
is 
\begin{equation}
\mathcal{L}=\mathcal{L}_{YM}+\mathcal{L}_{CP}+\mathcal{L}_{\mathrm{matter}},
\end{equation}
 where
\begin{equation}
\mathcal{L}_{YM}=\sum_{a,b}x_{ab}\left(H^{a}\right)^{\mu\nu}(H^{b})^{\rho\lambda}\mbox{Tr}\left(F_{\mu\rho}F_{\nu\lambda}\right),\label{eq:L gauge}
\end{equation}
\begin{equation}
\mathcal{L}_{CP}=\sum_{a,b}y_{ab}\left(H^{a}\right)^{\mu\nu}(H^{b})^{\rho\lambda}\mbox{Tr}\left(F_{\mu\rho}\tilde{F}_{\nu\lambda}\right),\label{eq:Lagrangian CP violating}
\end{equation}
\begin{equation}
\mathcal{L}_{\mathrm{matter}}=\sum_{a}z_{a}\left(H^{a}\right)^{\mu\nu}D_{\mu}\phi^{\dagger}D_{\nu}\phi-V_{\phi}(\phi,\phi^{\dagger}),\label{eq:L matter}
\end{equation}
for some constants $x_{ab},$$y_{ab}$, and $z_{a}$, $\tilde{F}_{\alpha\beta}\equiv\frac{1}{2}\epsilon_{\alpha\beta\mu\nu}F^{\mu\nu}$
is the dual field strength tensor, and $V_{\phi}(\phi,\phi^{\dagger})$
is a gauge invariant potential. The residual basis dependence of $\mathcal{L}$
is removed by applying the permutation symmetry Eq.~(\ref{eq:perm sym frame ind proof});
this leads to the restriction
\begin{equation}
x_{ab}=\begin{cases}
x+x' & a=b\\
x & a\neq b
\end{cases}\qquad y_{ab}=\begin{cases}
y+y' & a=b\\
y & a\neq b
\end{cases}\qquad z_{a}=z\label{eq:restrictions due to perm sym}
\end{equation}
for some constants $x$, $x'$, $y$,$y$', and $z$. The Lagrangian
then reduces to 
\begin{eqnarray}
\mathcal{L} & = & x\sum_{a,b}\left(H^{a}\right)^{\mu\nu}(H^{b})^{\rho\lambda}\mbox{Tr}\left(F_{\mu\rho}F_{\nu\lambda}\right)+x'\sum_{a}\left(H^{a}\right)^{\mu\nu}(H^{a})^{\rho\lambda}\mbox{Tr}\left(F_{\mu\rho}F_{\nu\lambda}\right)\nonumber \\
 &  & +y\sum_{a,b}\left(H^{a}\right)^{\mu\nu}(H^{b})^{\rho\lambda}\mbox{Tr}\left(F_{\mu\rho}\tilde{F}_{\nu\lambda}\right)+y'\sum_{a}\left(H^{a}\right)^{\mu\nu}(H^{a})^{\rho\lambda}\mbox{Tr}\left(F_{\mu\rho}\tilde{F}_{\nu\lambda}\right)\nonumber \\
 &  & +z\sum_{a}\left(H^{a}\right)^{\mu\nu}D_{\mu}\phi^{\dagger}D_{\nu}\phi-V_{\phi}(\phi,\phi^{\dagger})\\
 & = & x\mbox{Tr}\left(F_{\mu\rho}F^{\mu\rho}\right)+y\mbox{Tr}\left(F_{\mu\rho}\tilde{F}^{\mu\rho}\right)+zD^{\mu}\phi^{\dagger}D_{\mu}\phi-V_{\phi}(\phi,\phi^{\dagger})\label{eq:YM + matter lagrangian from BTGT}
\end{eqnarray}
which is frame independent. The frame dependent $x'$ and $y'$ terms
are zero because $F_{\mu\nu}$ and its dual are anti-symmetric while
the coefficients satisfy the identity\lyxdeleted{danielchung}{Mon Dec 30 08:25:17 2019}{
} $\left(H^{a}\right)^{\mu\nu}(H^{a})^{\rho\lambda}=\left(H^{a}\right)^{\mu\rho}(H^{a})^{\nu\lambda}$
. Eq.~(\ref{eq:YM + matter lagrangian from BTGT}) with $x=-\frac{1}{4g^{2}}$,
$y=-\frac{\theta_{QCD}}{32\pi^{2}}$ and $z=1$ corresponds to the
usual Yang-Mills plus matter Lagrangian. We see from Eq.~(\ref{eq:YM + matter lagrangian from BTGT})
that the only $\psi_{(a)}^{\mu}$ dependence of $S[\psi_{(a)}^{\mu},U_{a},\phi]$
occurs in $D^{\mu}\phi$ and $F^{\mu\nu}$, and from Eq.~(\ref{eq:F in terms of vector})
and Eq.~(\ref{eq:Dphi in terms of vector}) we see that this $\psi_{(a)}^{\mu}$
dependence only occurs in form of Eq.~(\ref{eq:V in terms of Ua}).
Therefore the action $S[\psi_{(a)}^{\mu},U_{a},\phi]$ can written
as Eq.~(\ref{eq:btgt action is vector gauge action}), where the
action $\tilde{S}[V^{\mu},\phi]$ is the usual non-Abelian gauge theory
action in terms of gauge connection $V^{\mu}$. Since $\tilde{S}$
contains no $\psi_{(a)}^{\mu}$ dependence, the BTGT action $S[\psi_{(a)}^{\mu},U_{a},\phi]$
is thus frame independent.

One possible limitation that prevents an extension of frame independence
to the quantum theory itself is the path integral measure. Consider
a change of variables of integration in the partition function from
$\theta_{a}$ to $V_{\mu}$ defined in Eq.\ (\ref{eq:V in terms of Ua}).
The change of variables affects the theory via the measure change
by
\begin{equation}
D\theta=\left|\frac{\delta\theta_{a}}{\delta V_{\mu}}\right|DV=\mathcal{J}_{\psi}DV
\end{equation}
 where the subscript on the Jacobian $J_{\psi}$ indicates that the
theory may still be frame dependent. For instance, in the Abelian
case we write the Jacobian inverse as 
\begin{equation}
\mathcal{J}_{\psi}^{-1}=\det_{\mu a}\left(\eta^{aa}\psi_{(a)}^{\mu}\psi_{(a)}^{\lambda}\partial_{\lambda}\right)=\det_{\mu a}\left((H^{a})_{\phantom{\mu}\lambda}^{\mu}\partial^{\lambda}\right),\label{eq:J explicit}
\end{equation}
and we see that explicit frame dependence coming from the zero modes
of derivative operator $(H^{a})_{\phantom{\mu}\lambda}^{\mu}\partial^{\lambda}$.
Nonetheless, this Jacobian should be frame independent apart from
the zero modes, which do not affect perturbation theory.\lyxdeleted{danielchung}{Mon Dec 30 08:25:17 2019}{
} This is consistent with the explicit one loop computations done
in both the Abelian \cite{Chung:2017zck} and non-Abelian \cite{Basso:2019yap}
cases. The residual $\psi_{(a)}^{\mu}$ dependence of the theory might
also be removable\lyxdeleted{danielchung}{Mon Dec 30 08:25:17 2019}{
} by averaging over all $\psi_{(a)}^{\mu}$, i.e. taking a path integral
over all possible $\psi_{(a)}^{\mu}$. Another possible (but even
less likely) obstruction to the theory being frame independent would
be the non-invariance of the measure under the symmetries such as
the $S_{4}$ permutation and parity symmetries. We will not address
these issues further in this paper.

There are also higher mass dimension non-renormalizable terms with
$\psi_{(a)}^{\mu}$ frame dependence that can be written down in the
Lagrangian\lyxdeleted{danielchung}{Mon Dec 30 08:25:17 2019}{ } consistent
with all the symmetries. . For example, terms such as
\begin{equation}
\mathcal{L}_{\mathrm{f.}\,\mathrm{dep}.\,1}=\sum_{a}\left(H^{a}\right)^{\mu\nu}\left(H^{a}\right)^{\rho\sigma}D_{\mu}\phi^{\dagger}D_{\nu}\phi D_{\rho}\phi^{\dagger}D_{\sigma}\phi\mbox{Tr}\left(F^{2}\right)\label{eq:L frame dep 1}
\end{equation}
and
\begin{align}
\mathcal{L}_{\mathrm{f.}\,\mathrm{dep}.\,2} & =\sum_{a}(H^{a})^{\mu\nu}(H^{a})^{\rho\sigma}(H^{a})^{\alpha\beta}(H^{a})^{\gamma\delta}F_{\mu\alpha}F_{\nu\beta}F_{\rho\gamma}F_{\sigma\delta}\label{eq:L frame dep 2}
\end{align}
belong in this category of operators. Even though they are not eliminated
by pure gauge basis invariance, they are forbidden by a scaling symmetry
of $H_{a}\rightarrow e^{\phi_{a}}H_{a}$, $U_{a}\rightarrow e^{-\phi_{a}}U_{a}$,
$U_{a}^{\dagger}\rightarrow e^{-\phi_{a}}U_{a}^{\dagger}$ for real
global parameter $\phi_{a}$. Dividing by $\mbox{Tr}[H_{a}]^{2}$
in the sum of Eq.~(\ref{eq:L frame dep 1}) would yield an invariant,
but these terms will not arise in the effective action as as counter-terms
do not involve inverse powers of fields. We have thus far ignored
the measure issue under this scaling, but we suspect that it should
not be an issue outside of zero modes. This is discussed further\lyxdeleted{danielchung}{Mon Dec 30 08:25:17 2019}{
} in Appendix \ref{sec:Ua Ha scaling and nonrenorm terms}.

\section{\label{sec:Conclusions}Summary}

In this paper, we have clarified the representational aspects of the
BTGT quantization variable $\theta_{a}^{A}H_{\mu\nu}^{a}$ which contains
the information about the gauge vierbein analog variable $\left[G_{(f)\,\beta}^{\alpha}(x)\right]^{i}$. 

Our first result was to show that $(H^{a})^{\mu\nu}$ when written
in terms of $\psi_{(a)}^{\mu}$ in Eq.~(\ref{eq:intermsofpsi}) can
be understood as a tensor made from a bilinear combination of Jacobian
factors (Eq.~(\ref{eq:defofglobalfield})) in going from an arbitrary
chosen inertial frame $x[1]$ to another inertial frame. This means
$\theta_{a}^{A}$ are frame $x[1]$ choice dependent fields. Afterwards,
we showed through Eq.~(\ref{eq:10exp}) how a manifestly covariant
description of $\left[G_{(f)\,\beta}^{\alpha}(x)\right]^{i}$ without
reference to $x[1]$ can be accomplished through $(G^{\alpha\beta})^{ij}$
which has $10D(\mathcal{G})$ degrees of freedom (where $D(\mathcal{G})$
is the dimension of the gauge group transforming according to Eq.~(\ref{eq:Gtransform-1})).
This means by quantizing using $\theta_{a}^{A}$, we are setting $6D(\mathcal{G})$
degrees of freedom to effectively to zero. 

Despite this arbitrariness, because BTGT was defined through a field
redefinition from $A_{\mu}^{B}$ to $\theta_{a}^{B}$ in \cite{Chung:2016lhv,Chung:2017zck,Basso:2019yap},
the BTGT formalism is manifestly Lorentz invariant and $x[1]$ choice
independent. As a corollary, we have shown through Eqs.~(\ref{eq:equiv1})
and (\ref{eq:equiv2}) that different $x[1]$ choices leads to an
equivalence class of $\theta_{a}^{A}$. This equivalence class is
distinct from the BTGT and gauge symmetry, and it is inherent in the
usage of $\theta_{a}^{A}$ for quantization. Although this last statement
is in some sense trivial since the equivalence arises from merely
a field redefinition, computations at one loop \cite{Chung:2016lhv,Chung:2017zck,Basso:2019yap}
did not show a sensitivity to the path integral variable change Jacobian.
This makes this equivalence class statement less trivial.

We then partially explained why the $x[1]$ frame dependence is disappearing
in \cite{Chung:2016lhv,Chung:2017zck,Basso:2019yap} by showing explicitly
in Sec.~\ref{sec:Frame-independence-without} that BTGT invariance,
gauge invariance, renormalizablity, and a couple of discrete symmetries
associated with the frame-dependent variables ($S_{4}$ permutation
symmetry and a parity symmetry related to $PT$ symmetry) make the
action $x[1]$ choice independent even when the action depends on
both $\psi_{(a)}^{\mu}$ and $U_{a}=\exp\left[i\theta_{a}^{A}T^{A}\right]$
(both of which are $x[1]$ dependent). The full quantum generating
functional is almost $x[1]$ independent except for the issues associated
with the zero modes of the Jacobian associated with the path integration
measure chage. Hence, the perturbative computation is also argued
to be $x[1]$ independent.

The work presented in this paper has several obvious extensions. First,
although we have focused in Sec.~\ref{sec:Frame-independence-without}
on renormalizable BTGT theories, the frame independence may be generalizable
to nonrenormalizable theories through extended symmetries of the form
in Eq.~(\ref{eq:Ha Ua scaling}). Secondly, we have explicitly constructed
the the frame independence in Sec.~\ref{sec:Frame-independence-without}
including only gauged scalar matter fields. It would be interesting
to extend this to higher spin fields. Thirdly, we noted in Eq.~(\ref{eq:F btgt formalism})
that the non-Abelian field strength tensor in the BTGT formalism is
homogeneous in the derivatives just as in the Abelian theory. It would
be interesting to use this property as well as the property $D_{(a)}U_{a}=0$
for constructing novel semiclassical solutions.

Perhaps the most interesting extension is to embed Eq.~(\ref{eq:10exp})
into coset model of gauge fields. (For other related efforts in this
direction, see e.g. \cite{Kovner:1992pu,Gaiotto:2014kfa,Hofman:2018lfz}.)
For this effort, it would be useful to rewrite the BTGT symmetry in
terms of $G_{\mu\nu}$ only without referring to the $U_{a}$ field.
Such coset constructions will naively generate $6D(\mathcal{G})$
additional degrees of freedom than what is observed. It would be interesting
whether these additional degrees of freedom can be sufficiently hidden
for phenomenological consistency.
\begin{acknowledgments}
This work was supported in part by the DOE through grant DE-SC0017647. 
\end{acknowledgments}

\appendix

\section{Definitions \label{sec:Definitions}}

In this section, we explicitly define some of the terms used in the
paper.

\subsection{\label{sub:Basis-fields}Basis fields $\psi_{(a)}^{\mu}$}

The basis fields $\psi_{(a)}^{\mu}$ are four constant/global fields
$\left\{ \psi_{(0)}^{\mu},\psi_{(1)}^{\mu},\psi_{(2)}^{\mu},\psi_{(3)}^{\mu}\right\} $
that form an orthogonal basis of spacetime that corresponds to a particular
rest frame $x[1]$: 
\begin{equation}
\psi_{(a)}^{\mu}\psi_{(b)\mu}=\delta_{ab}\psi_{(a)}^{\mu}\psi_{(a)\mu},\label{eq:psi orthogonal}
\end{equation}

\begin{equation}
\sum_{a=0}^{3}\frac{\psi_{(a)}^{\mu}\psi_{(a)}^{\nu}}{\psi_{(a)}^{\lambda}\psi_{(a)\lambda}}=\eta^{\mu\nu},\label{eq:psi completeness}
\end{equation}
where $\eta^{\mu\nu}$ is the spacetime Lorentzian metric. Typically,
one chooses a scaling for $\psi_{(a)}^{\mu}$ such that 
\begin{equation}
\left(\psi_{(a)}^{\lambda}\psi_{(a)\lambda}\right)^{-1}=\eta^{aa}=\left\{ +1,-1,-1,-1\right\} ,
\end{equation}
In which case, the properties of Eqs.~(\ref{eq:psi orthogonal})
and (\ref{eq:psi completeness}) can be expressed as
\begin{equation}
\psi_{(a)}^{\mu}\psi_{(b)\mu}=\eta_{ab}\qquad\mbox{and}\qquad\sum_{a=0}^{3}\eta^{aa}\psi_{(a)}^{\mu}\psi_{(a)}^{\nu}=\eta^{\mu\nu}.
\end{equation}
See section \ref{sub:Rescaling-symmetry} for more details on the
scaling symmetry of $\psi_{(a)}^{\mu}$. Note that $a$ is a fictitious
Lorentz index, while $\mu$ is a real spacetime tensor index.

\subsection{Frame independence \label{sub:def of no preferred basis}}

Let $\psi_{(a)}^{\mu}$ be the\lyxdeleted{danielchung}{Mon Dec 30 08:25:17 2019}{
} basis defined in Sec.~\ref{sub:Basis-fields} and $\phi=\left\{ \phi_{1},\phi_{2},\dots\right\} $
be a set of usual local fields with a manifestly covariant Lorentz
tensor representation. An\textbf{ }action\textbf{ $S\left[\psi,\phi\right]$
}is frame independent if there exists a change of variables such that
\begin{equation}
S\left[\psi,\phi\right]=\tilde{S}\left[\Phi\right]
\end{equation}
for some new action $\tilde{S}\left[\Phi\right]$ that is independent
of $\psi$ (where $\Phi$ is a new set of local fields with a manifestly
covariant Lorentz tensor representation). A theory $Z$ with action
$S[\psi,\phi]$ is frame independent if there exists some change of
variables\lyxdeleted{danielchung}{Mon Dec 30 08:25:17 2019}{ } such
that 
\begin{equation}
Z=\int D\phi e^{iS\left[\psi,\phi\right]}\propto\int D\Phi e^{i\tilde{S}\left[\Phi\right]}
\end{equation}
for some new action $\tilde{S}\left[\Phi\right]$ and measure $D\Phi$
that are independent of $\psi$.

\subsection{Lorentz transformations}

The BTGT index $a$ is a label and not Lorentz tensor index. Therefore
the BTGT field $\theta_{a}$ transforms as a scalar and the basis
$\psi_{(a)}^{\mu}$ as vectors. Under a Lorentz transformation $x\rightarrow\Lambda x$,
\begin{align}
\psi_{(a)}^{\mu} & \rightarrow\Lambda_{\phantom{\mu}\nu}^{\mu}\psi_{(a)}^{\nu}\nonumber \\
\theta_{a}\left(x\right) & \rightarrow\theta_{a}\left(\Lambda^{-1}x\right)\nonumber \\
\phi\left(x\right) & \rightarrow\phi\left(\Lambda^{-1}x\right)\label{eq:field lorentz transforms}
\end{align}
where $\phi\left(x\right)$ is some scalar matter field. The action
must be invariant under the Lorentz transformations of Eq.~(\ref{eq:field lorentz transforms}).

\subsection{Gauge transformations}

Given a matter fields $\phi$ transforming as the fundamental representation
of the gauge group $\mathcal{G}$, the BTGT fields $U_{a}$ are defined
to transform under a gauge transformation such that $U_{a}^{-1}\phi$
and $\phi^{\dagger}U_{a}$ are gauge group singlets. Therefore the
gauge transformations of $U_{a}$ are 
\begin{align}
U_{a}\left(x\right) & \overset{\mathrm{gauge}}{\rightarrow}e^{i\Gamma\left(x\right)}U_{a}\left(x\right),\\
U_{a}^{-1}\left(x\right) & \overset{\mathrm{gauge}}{\rightarrow}U_{a}^{-1}\left(x\right)e^{-i\Gamma\left(x\right)},
\end{align}
where $U_{a}\left(x\right)=e^{iT^{A}\theta_{a}^{A}\left(x\right)}$.

\subsection{BTGT transformations}

Given a set of four scalar BTGT fields $U_{a}$, the BTGT transformation
is given by 
\begin{equation}
U_{a}\left(x\right)\stackrel{\mathrm{btgt}}{\rightarrow}U_{a}\left(x\right)e^{iZ_{a}\left(x\right)},\label{eq:BTGT symmetry defintion}
\end{equation}
where $Z_{a}(x)=T^{A}Z_{a}^{A}\left(x\right)$ satisfies the zero
mode constraint 
\begin{equation}
\psi_{(a)}^{\mu}\partial_{\mu}Z_{a}\left(x\right)=0.\qquad\left(\mbox{no sum over \ensuremath{a}}\right)\label{eq:Zero mode constraint}
\end{equation}
The BTGT variation is defined to be zero for all other fields: 
\begin{equation}
\delta_{\mathrm{btgt}}\psi_{(a)}^{\mu}=\delta_{\mathrm{btgt}}\phi=0.
\end{equation}

\subsection{Permutation symmetry of BTGT labels}

Relabeling symmetry is generated by the exchanges

\begin{equation}
\left\{ H^{a},U_{a}\right\} \leftrightarrow\left\{ H^{b},U_{b}\right\} \label{eq:relabel symmetry H and U}
\end{equation}
for any $a$ and $b$. In terms of $\psi_{(a)}$ and $U_{a}$ it is
\begin{equation}
\left\{ \eta^{aa},\psi_{(a)},U_{a}\right\} \leftrightarrow\left\{ \eta^{bb},\psi_{(b)},U_{b}\right\} \label{eq:relabel symmetry eta psi U}
\end{equation}
We impose this condition on BTGT to obtain ordinary gauge theory.
One possible complication of this permutation symmetry is that it
affects the interpretation of $\psi_{(a)}^{\mu}$ as the Jacobian
of a frame change from $x[1]$ to $x[2]$. The swapping of Latin index
$a$ on $\eta_{aa}$ must be matched with some swapping of spacetime
indices on $\eta_{\mu\nu}$. When both Latin and Greek indices are
free and not summed seems to be the only possible problematic case.
However, there was no case found for which this would occur. The issue
does not therefore seem important at this time.

\subsection{Rescaling symmetry of $\psi_{(a)}^{\mu}$ \label{sub:Rescaling-symmetry}}

Each $\psi_{(a)}^{\mu}$ has an independent rescaling symmetry
\begin{equation}
\psi_{(a)}^{\mu}\rightarrow\lambda_{a}\psi_{(a)}^{\mu}\label{eq:rescaling symmetry of psi}
\end{equation}
for any constant $\lambda_{a}\neq0$. Note that the properties of
$\psi_{(a)}^{\mu}$ given by Eqs.~(\ref{eq:psi orthogonal}) and
(\ref{eq:psi completeness}) are invariant under the transformation
of Eq.~(\ref{eq:rescaling symmetry of psi}). One can normalize the
basis fields $\psi_{(a)}^{\mu}$ such that
\begin{equation}
\psi_{(a)}^{\lambda}\psi_{(a)\lambda}=\eta_{aa}=\left\{ +1,-1,-1,-1\right\} \label{eq:app psi normalized choice}
\end{equation}
by using a rescaling of $\lambda_{a}=1/\sqrt{\left|\psi_{(a)}^{\lambda}\psi_{(a)\lambda}\right|}$.
In that case, there remains a residual parity symmetry of the form
\begin{equation}
\psi_{(a)}^{\mu}\rightarrow\mbox{sign}\left(\lambda_{a}\right)\psi_{(a)}^{\mu}\label{eq:psi parity sym}
\end{equation}
that corresponds to the $PT$ discrete subgroup of the $O(1,3)$ Lorentz
group.

\subsection{Pure gauge basis invariance}

A pure gauge configuration is when $U_{a}=e^{i\Gamma}$, which is
gauge equivalent to the identity element\textbf{ $\mathbf{1}\in\mathcal{G}$.}
When $U_{a}$ is pure gauge, the theory is invariant under a basis
change of the form
\begin{equation}
\psi_{(a)}^{\mu}\rightarrow\sum_{b}\Lambda_{\phantom{b}a}^{b}\psi_{(b)}^{\mu}
\end{equation}
where $\Lambda_{\phantom{b}a}^{b}\in O(1,3)$ is some Lorentz transformation
and basis set has been normalized via Eq.~(\ref{eq:app psi normalized choice}).

\section{The only $U_{a}$ dependence of $\mathcal{L}$ is $U_{a}\partial_{(a)}U_{a}^{-1}$
and $\partial_{\mu}\left(U_{a}\partial_{(a)}U_{a}^{-1}\right)$ \label{sub:Lemma of Ua dependence}}

In this section we will show that if the action $S=S\left[\psi_{(a)}^{\mu},U_{a},\phi\right]$
is invariant under the BTGT symmetry defined in Eq.~(\ref{eq:BTGT symmetry defintion}),
then the field dependence of the Lagrangian of that action up to two
derivatives of the fields can be expressed as 
\begin{equation}
\mathcal{L}=\mathcal{L}\left(\psi_{(a)}^{\mu},U_{a}\partial_{(a)}U_{a}^{-1},\partial_{\mu}\left(U_{a}\partial_{(a)}U_{a}^{-1}\right),\phi,\partial_{\mu}\phi\right)\label{eq:Ua dependence lemma result}
\end{equation}
where $\partial_{(a)}U_{a}^{-1}=\psi_{(a)}^{\mu}\partial_{\mu}U_{a}^{-1}$. 

Let's begin. In terms of all the fields and their derivatives, the
most general Lagrangian has field dependence of the form
\begin{equation}
\mathcal{L}=\mathcal{L}\left(\psi_{(a)}^{\mu},U_{a},\partial_{\mu}U_{a},\partial_{\mu}\partial_{\nu}U_{a},U_{a}^{-1},\partial_{\mu}U_{a}^{-1},\partial_{\mu}\partial_{\nu}U_{a}^{-1},\phi,\partial_{\mu}\phi\right)\label{eq:L explicit 1}
\end{equation}
where the space-time dependence is implicit. Since $\partial_{\mu}U_{a}=-U_{a}\partial_{\mu}\left(U_{a}^{-1}\right)U_{a}$,
the Lagrangian has field dependence
\begin{equation}
\mathcal{L}\left(x\right)=\mathcal{L}\left(\psi_{(a)}^{\mu},U_{a},U_{a}^{-1},\partial_{\mu}U_{a}^{-1},\partial_{\mu}\partial_{\nu}U_{a}^{-1},\phi,\partial_{\mu}\phi\right),\label{eq:L field dependence 2}
\end{equation}
without loss of generality compared to Eq.\,(\ref{eq:L explicit 1}).

To obtain BTGT singlets in $\mathcal{L}$, one must have pairs of
$U_{a}$ and some derivative of $U_{a}^{-1}$, with $U_{a}$ appearing
to the left of $U_{a}^{-1}$. These $U_{a}$, $U_{a}^{-1}$ pairs
are the building blocks of making larger BTGT invariants. We can show
by exhaustion that each of them can be expressed in terms of $U_{a}\partial_{(a)}U_{a}^{-1}$
and its derivatives. Let us list all such building blocks up to two
derivatives acting on $U_{a}$. There are the field independent invariant
combinations 
\begin{eqnarray}
\mathbf{1} & = & U_{a}U_{a}^{-1}=U_{a}^{-1}U_{a},\\
\epsilon_{k_{1}k_{2}\cdots k_{N}} & = & \epsilon_{i_{1}i_{2}\cdots i_{N}}\left(U_{a}\right)_{i_{1}k_{1}}\left(U_{a}\right)_{i_{2}k_{2}}\cdots\left(U_{a}\right)_{i_{N}k_{N}},
\end{eqnarray}
and the non-trivial invariant 
\begin{equation}
U_{a}\partial_{(a)}\left(U_{a}^{-1}\right)
\end{equation}
 and its derivative $\partial_{\mu}\left(U_{a}\partial_{(a)}U_{a}^{-1}\right).$
The other invariants are $\partial_{(a)}\left(U_{a}\right)U_{a}^{-1}$,
$U_{a}\partial_{(a)}\left(\partial_{(a)}\left(U_{a}^{-1}\right)\right)$,
$\partial_{(a)}\left(U_{a}\right)\partial_{(a)}\left(U_{a}^{-1}\right)$,
and $\partial_{(a)}\left(\partial_{(a)}\left(U_{a}\right)\right)U_{a}^{-1}$.
Each of these can be expressed in terms of $U_{a}\partial_{(a)}U_{a}^{-1}$
and its derivative:
\begin{align}
\partial_{(a)}\left(U_{a}\right)U_{a}^{-1} & =-U_{a}\partial_{(a)}\left(U_{a}^{-1}\right)\\
U_{a}\partial_{(a)}\left(\partial_{(a)}\left(U_{a}^{-1}\right)\right) & =\partial_{(a)}\left(U_{a}\partial_{(a)}\left(U_{a}^{-1}\right)\right)+U_{a}\partial_{(a)}\left(U_{a}^{-1}\right)U_{a}\partial_{(a)}\left(U_{a}^{-1}\right)\\
\partial_{(a)}\left(U_{a}\right)\partial_{(a)}\left(U_{a}^{-1}\right) & =-U_{a}\partial_{(a)}\left(U_{a}^{-1}\right)U_{a}\partial_{(a)}\left(U_{a}^{-1}\right)\\
\partial_{(a)}\left(\partial_{(a)}\left(U_{a}\right)\right)U_{a}^{-1} & =-\partial_{(a)}\left(U_{a}\partial_{(a)}\left(U_{a}^{-1}\right)\right)+U_{a}\partial_{(a)}\left(U_{a}^{-1}\right)U_{a}\partial_{(a)}\left(U_{a}^{-1}\right)
\end{align}
BTGT invariants are made from these building blocks and can therefore
be expressed in terms of $U_{a}\partial_{(a)}U_{a}^{-1}$ and its
derivative.

In addition, we have the following combinations that transform as
BTGT adjoints $U_{a}^{-1}\partial_{(a)}\left(U_{a}\right)$, $\partial_{(a)}\left(U_{a}^{-1}\right)U_{a}$,
$\partial_{(a)}\left(\partial_{(a)}\left(U_{a}^{-1}\right)\right)U_{a}$,
$\partial_{(a)}\left(U_{a}^{-1}\right)\partial_{(a)}\left(U_{a}\right)$,
and $U_{a}^{-1}\partial_{(a)}\left(\partial_{(a)}\left(U_{a}\right)\right)$.
These can be expressed as 
\begin{align}
U_{a}^{-1}\partial_{(a)}\left(U_{a}\right) & =U_{a}^{-1}\left(\partial_{(a)}\left(U_{a}\right)U_{a}^{-1}\right)U_{a}\\
\partial_{(a)}\left(U_{a}^{-1}\right)U_{a} & =U_{a}^{-1}\left(U_{a}\partial_{(a)}\left(U_{a}^{-1}\right)\right)U_{a}\\
\partial_{(a)}\left(\partial_{(a)}\left(U_{a}^{-1}\right)\right)U_{a} & =U_{a}^{-1}\left(U_{a}\partial_{(a)}\left(\partial_{(a)}\left(U_{a}^{-1}\right)\right)\right)U_{a}\\
\partial_{(a)}\left(U_{a}^{-1}\right)\partial_{(a)}\left(U_{a}\right) & =-U_{a}^{-1}\left(U_{a}\partial_{(a)}\left(U_{a}^{-1}\right)\right)\left(U_{a}\partial_{(a)}U_{a}^{-1}\right)U_{a}\\
U_{a}^{-1}\partial_{(a)}\left(\partial_{(a)}\left(U_{a}\right)\right) & =U_{a}^{-1}\left(\partial_{(a)}\left(\partial_{(a)}\left(U_{a}\right)\right)U_{a}^{-1}\right)U_{a}
\end{align}
All of these BTGT adjoints have the form of 
\begin{equation}
\mbox{BTGT adjoint}=U_{a}^{-1}\left(\mbox{BTGT invariant}\right)U_{a}
\end{equation}
Since $\mathcal{L}$ must be a BTGT invariant, the BTGT adjoint objects
must eventually appear in a group trace composed only of BTGT adjoint
objects \cite{Mountain:1998}. Since $U_{a}$ is the only field with
a BTGT charge, these are the only BTGT adjoint objects that appear
in $\mathcal{L}$. Therefore, they can only appear with each other
in a trace. Inside the trace, they become equivalent to BTGT invariants:
\begin{align}
\mbox{Tr}\left[\left(\mbox{BTGT adj.}\right)\cdots\left(\mbox{BTGT adj.}\right)\right] & =\mbox{Tr}\left[U_{a}^{-1}\left(\mbox{BTGT inv.}\right)U_{a}\cdots U_{a}^{-1}\left(\mbox{BTGT inv.}\right)U_{a}\right]\\
 & =\mbox{Tr}\left[\left(\mbox{BTGT inv.}\right)\cdots\left(\mbox{BTGT inv.}\right)\right]
\end{align}
As a result, the only non-trivial $U_{a}$ dependence of $\mathcal{L}$
is on $U_{a}\partial_{(a)}\left(U_{a}^{-1}\right)$ and its derivative
$\partial_{\mu}U_{a}\partial_{(a)}\left(U_{a}^{-1}\right)$.

\section{Frame independence for non-renormalizable terms \label{sec:Ua Ha scaling and nonrenorm terms}}

When vierbein field $G_{\phantom{\mu}\nu}^{\mu}$ is parameterized
as Eq.~(\ref{eq:btgtfield}) in terms of $H^{a}$ and $U_{a}$, there
are certain transformations of $H^{a}$ and $U_{a}$ that keep $G_{\phantom{\mu}\nu}^{\mu}$
invariant. These symmetries include the relabeling permutation symmetry
of Eq.~(\ref{eq:relabel symmetry H and U}), and the $\psi_{(a)}^{\mu}$
scaling/parity symmetry of Eqs.~(\ref{eq:rescaling symmetry of psi})
and (\ref{eq:psi parity sym}), and the pure gauge basis invariance
defined in Appendix \ref{sec:Definitions}. Another symmetry that
would keep Eq.~(\ref{eq:btgtfield}) invariant is 
\begin{equation}
H^{a}\rightarrow e^{\phi_{a}}H^{a},\quad U_{a}\rightarrow e^{-\phi_{a}}U_{a},\quad U_{a}^{\dagger}\rightarrow e^{-\phi_{a}}U_{a}^{\dagger},\label{eq:Ha Ua scaling}
\end{equation}
where $\phi_{a}$ is a real global parameter. This scaling symmetry
is designed so that $H^{a}$ appears only in the form of $H^{a}U_{a}$
or $H^{a}U_{a}^{\dagger}$ (no sum over $a$).

The symmetry of Eq.~(\ref{eq:Ha Ua scaling}) prevents frame dependent
terms like Eqs.~(\ref{eq:L frame dep 1}) and (\ref{eq:L frame dep 2})
from emerging in any effective action potential. However, the scaling
also corresponds to a deformation of the properties of both $H^{a}$
and $U_{a}$ as previously defined. For example, the inverse to $U_{a}$
would satisfy
\begin{equation}
U_{a}^{-1}=\mbox{Tr}(H^{a})^{2}U_{a}^{\dagger}\label{eq:Ua inv redef}
\end{equation}
instead of $U_{a}^{-1}=U_{a}^{\dagger}$. In addition, $\theta_{a}=\theta_{a}^{A}T^{A}$
is deformed from the Lie algebra under the scaling of Eq.~(\ref{eq:Ha Ua scaling})
by 
\begin{equation}
\theta_{a}^{A}T^{A}\rightarrow\theta_{a}^{A}T^{A}+i\phi_{a}\mathbf{1},\label{eq:theta scaling Abelian}
\end{equation}
where \textbf{$\mathbf{1}$} is multiplicative algebra identity such
that $\left[T^{A},\mathbf{1}\right]=0$. The imaginary component of
Eq.~(\ref{eq:theta scaling Abelian}) is an analytic continuation
of the field $\theta_{a}$. Covariant derivatives would be modified
to 
\begin{equation}
D_{\mu}(\boldsymbol{\cdot})=\sum_{a}\mbox{Tr}(H^{a})U_{a}(H^{a})_{\phantom{\mu}\mu}^{\lambda}\partial_{\lambda}(U_{a}^{\dagger}\boldsymbol{\cdot}),\label{eq:covariant derivative HaUa scaling}
\end{equation}
which is obtained by replacing $U_{a}^{-1}$ with Eq.~(\ref{eq:Ua inv redef})
and $H^{a}$ with $H^{a}/\mbox{Tr}(H^{a})$. The same replacements
should be consistently done in the BTGT and gauge invariant Lagrangian.

\bibliographystyle{JHEP2}
\bibliography{nonabelian-btgtpaper}

\end{document}